\documentclass[12pt]{article}
\usepackage[utf8]{inputenc}
\usepackage[T1]{fontenc}
\usepackage{amsmath,amssymb,amsthm,commath,enumitem,mathtools}
\usepackage{bm}
\usepackage{kbordermatrix}
\usepackage[margin=2.54cm]{geometry}
\usepackage[normalem]{ulem}
\usepackage[numbers]{natbib}
\usepackage[normalem]{ulem}
\usepackage{graphicx}
\usepackage{booktabs}
\usepackage{caption}
\usepackage{color} 
\usepackage{url} 
\usepackage[small]{titlesec}
\usepackage{enumitem}
\usepackage[ruled,vlined]{algorithm2e}
\usepackage[tableposition=top,labelfont=bf]{caption}
\usepackage{tcolorbox}
\usepackage{layouts}
\usepackage{xcolor}
\usepackage{multirow}
\usepackage{makecell}
\graphicspath{ {./figures/} }
\usepackage[superscript,biblabel]{cite}
\usepackage{fancyhdr}
\usepackage{rotating}
\usepackage{lscape}
\usepackage{titletoc}

\allowdisplaybreaks
\let\originalleft\left
\let\originalright\right
\renewcommand{\left}{\mathopen{}\mathclose\bgroup\originalleft}
\renewcommand{\right}{\aftergroup\egroup\originalright}
\makeatletter
\newcommand*\bigcdot{\mathpalette\bigcdot@{.5}}
\newcommand*\bigcdot@[2]{\mathbin{\vcenter{\hbox{\scalebox{#2}{$\m@th#1\bullet$}}}}}
\makeatother

\title{%
   A Comparison Between Markov Switching Zero-inflated and Hurdle Models for Spatio-temporal Infectious Disease Counts\\[5pt]
  }
\author{Mingchi Xu, Dirk Douwes-Schultz\footnote{{{\it Corresponding author}: Dirk Douwes-Schultz, Department of Epidemiology, Biostatistics and Occupational Health, McGill University, 2001 McGill College Avenue, Suite 1200, Montreal, QC, Canada, H3A 1G1. {\it E-mail}: {\tt
				dirk.douwes-schultz@mail.mcgill.ca}.}} \hspace{1mm} and Alexandra M. Schmidt \\
				\textit{Department of Epidemiology, Biostatistics and Occupational Health} \\ \textit{McGill University, Canada }}
				
\date{\today}

\begin{document}

\maketitle

\begin{abstract}
In epidemiological studies, zero-inflated and hurdle models are commonly used to handle excess zeros in reported infectious disease cases. However, they can not model the persistence (changing from presence to presence) and reemergence (changing from absence to presence) of a disease separately. Covariates can sometimes have different effects on the reemergence and persistence of a disease. Recently, a zero-inflated Markov switching negative binomial model was proposed to accommodate this issue. We introduce a Markov switching negative binomial hurdle model as a competitor of that approach, as hurdle models are often also used as alternatives to zero-inflated models for accommodating excess zeroes. We begin the comparison by inspecting the underlying assumptions made by both models. Hurdle models assume perfect detection of the disease cases while zero-inflated models implicitly assume the case counts can be under-reported, thus we investigate when a negative binomial distribution can approximate the true distribution of reported counts. A comparison of the fit of the two types of Markov switching models is undertaken on chikungunya cases across the neighborhoods of Rio de Janeiro. We find that, among the fitted models, the Markov switching negative binomial zero-inflated model produces the best predictions and both Markov switching models produce remarkably better predictions than more traditional negative binomial hurdle and zero-inflated models. 

{\bf Keywords :} Bayesian inference;  Endemic-epidemic model; Under-reporting; Chikungunya    
\end{abstract}

\section{Introduction}

In epidemiological studies, disease counts taken at different spatial locations across different instants in time often contain a great number of zeros. In this case, a count distribution, like the Poisson or negative binomial distribution, is often unable to capture the large number of observed zero counts present in the data. Zero-inflated (ZI) and hurdle models \cite{RN7,RN8} are the two primary types of models that have been proposed to deal with count data with excess zeros. 

The first paper on ZI Poisson (ZIP) regression models handled count data with excess zeros by mixing a Poisson distribution and a distribution with a point mass at zero. \cite{RN6} In practice, due to the need for model flexibility, we can mix count distributions other than the Poisson with a distribution that has point mass at zero, like a ZI negative binomial model (ZINB) \cite{RN46}. Overall, we will refer to these models as ZI count (ZIC) models \cite{RN2}. In an epidemiological application of a ZIC model, a Bernoulli process is used to determine whether the disease is present \cite{RN49}. A one from the Bernoulli process indicates the disease is present and the number of cases comes from the count process, while a zero indicates the disease is absent and the number of cases is zero. Zeros can come from the zero mass process or the count process. Correspondingly, zero counts produced by a ZI model are often distinguished by "structural zeros", from the zero mass process, that correspond to the absence of the disease, and "sampling zeros", produced by the count process, which imply unreported cases from the at-risk population during the study period \cite{RN1}. We can also relate associated factors to the Bernoulli process that controls the presence/absence of the disease \cite{RN9}.

In comparison with ZI models, hurdle models also consist of two mixed parts: one is a zero-generating process, but the other part is a zero truncated count process. For example, a zero-truncated negative binomial distribution, which leads to a negative binomial hurdle model (NBH) \cite{RN50}. We can associate certain factors with the probability of disease presence in the same way as with ZI models \cite{RN12}. However, unlike ZI models, in hurdle models zeros cannot be produced by the at-risk population. Namely, in hurdle models, all zero counts are "structural zeros"  by construction. Therefore, compared to ZI models, a zero in a hurdle model, within an epidemiological context, can only arise due to the actual absence of the disease rather than the disease going undetected. Implicitly, this means that the disease is perfectly detected or that undetected cases are too few to be relevant, which is the main difference in the interpretation of zeros between ZI and hurdle models. 

Under-reporting is another challenge for researchers in epidemiology, where the reported disease counts can be less than the true counts. A zero-truncated count process, like a zero-truncated negative binomial distribution, can be applied to the reported counts when the disease is present under the perfect detection assumption of a hurdle model. However, a count distribution, such as the Poisson or negative binomial, can fail to approximate the true distribution of reported cases when the disease is present under the imperfect detection assumption of a ZI model. In Section~\ref{Section: NB approximation} we explore when the approximation can be acceptable.

Under the framework of spatio-temporal data, we can separate the presence of the disease into two categories: persistence (changing from presence to presence) and reemergence (changing from absence to presence). ZI models can only accommodate the characteristics of overall disease presence and cannot model reemergence and persistence separately. Sometimes, covariate effects can be quite different between the reemergence and persistence of an infectious disease \cite{RN5}. Recently, Douwes-Schultz and Schmidt (2022) \cite{RN5} extended the finite mixture ZIC model to a zero-state coupled Markov switching negative binomial model (ZS-CMSNB). They assumed the disease switched between periods of presence and absence in each area through a series of coupled Markov chains, where the reemergence and persistence were modelled separately \cite{RN5}. As a counterpart to the ZI models, in our framework, we follow the structure of hurdle models to assume that the zero mass process represents the reported cases when the disease is absent and a truncated count distribution (e.g. a zero truncated negative binomial distribution) represents the reported cases when the disease is present. We then assume a non-homogeneous Markov chain in each area switches the disease between the presence and absence states. We compare the Markov-switching negative binomial hurdle model to its zero-inflated counterpart on the fit, plausibility of assumptions and interpretation when modelling weekly reported chikungunya cases in Rio de Janeiro.

\subsection{Motivating example: Chikungunya cases in Rio de Janeiro }\label{sec:chik}

Chikungunya is an infectious disease that became endemic in Rio de Janeiro, Brazil in 2016 \cite{RN30}. For our study, we obtained publicly available data from the website of the Municipal Health Secretariat of Rio de Janeiro. The data comprises weekly counts across the 160 administrative districts of Rio de Janeiro. The data spans the period between January 2015 and May 2022. It is suspected that chikungunya started circulating unnoticed in the city before the first reported transmission \cite{RN31}. Due to a lack of social index information, we decided to exclude one small district, Paquet\'a Island.

\begin{figure}[!t]
 	\centering
  \includegraphics[width=\textwidth]{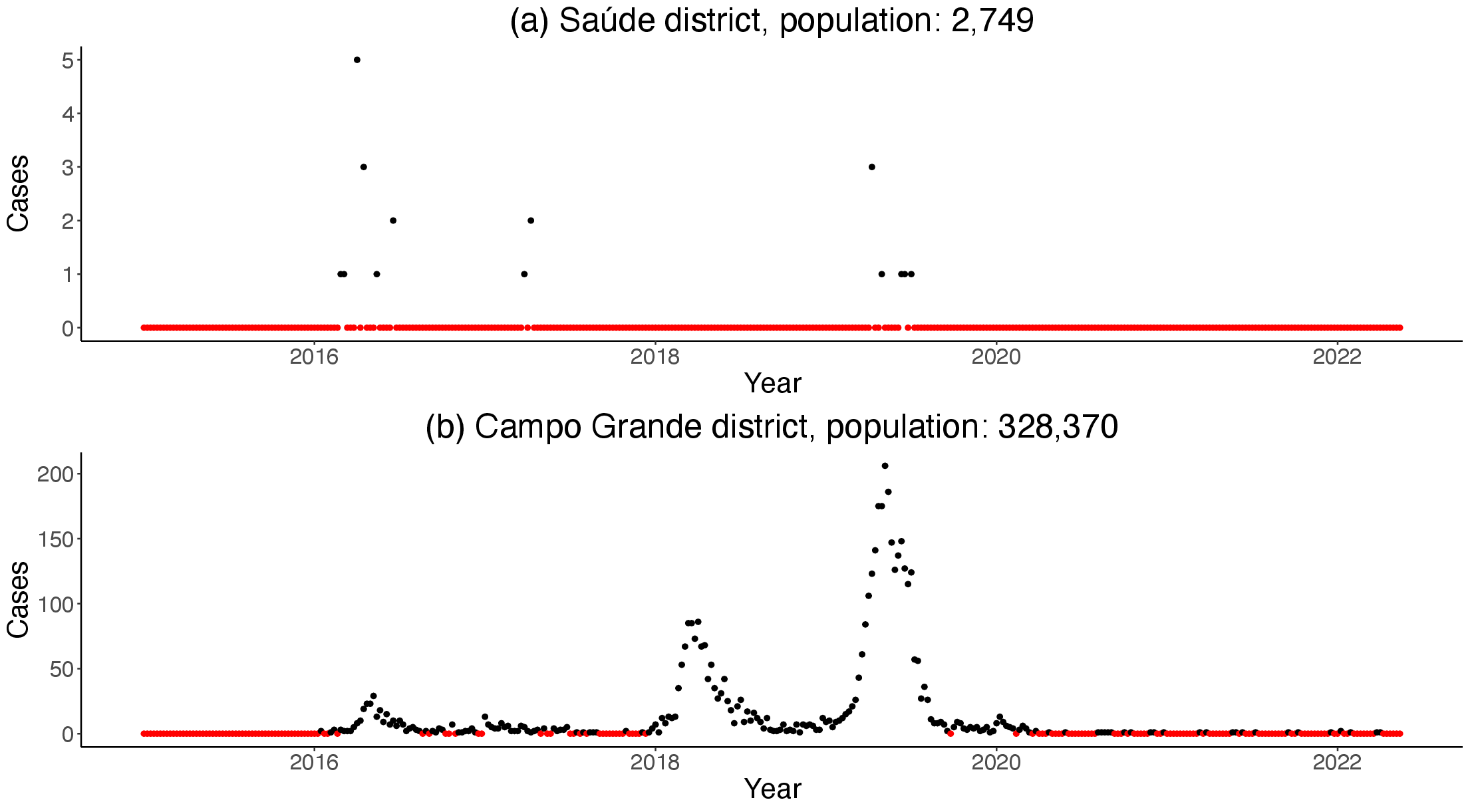}
	\caption{Weekly chikungunya infectious cases for (a) Sa\'ude district and (b) Campo Grande district; red solid circles represent 0 cases at the reported week during the study period.
 \label{fig: cases}} 
\end{figure}

Figure \ref{fig: cases} illustrates weekly chikungunya cases for two administrative districts of Rio de Janeiro, one with a small population (Sa\'ude) and the other one (Campo Grande) relatively large. In the Sa\'ude district, chikungunya cases were only reported for a couple of weeks and in most weeks no cases were reported (96.61\% of the study period). In the Campo Grande district, the disease showed a longer time of observed persistence (54.95\% of the study period) and is observed to reemerge (go from absence to presence) quicker. These differences in chikungunya persistence and reemergence probabilities at the district level could be explained by population differences, as there is a well-known inverse relationship between population and the rate of disease extinction in epidemiology \cite{RN27}. Socioeconomic factors may also partly explain the distinct patterns \cite{RN45}, since districts with lower Human Development Index (HDI) tend to lack tap water supply, which allows mosquitoes to breed in water storage containers and transmit disease \cite{RN35}. Because the mosquitoes are accustomed to urban utilities, the level of which is inversely correlated with the proportion of green areas (areas with agriculture, swamps and shoals, tree and shrub cover, and woody-grass cover) \cite{RN47}, the level of green area in a district could be inversely correlated with the disease transmission there. In this motivating example, we are mainly interested in investigating associations between certain factors, such as HDI and green areas, and Zika emergence and persistence, as well as the problem of future case prediction in a district to help policymakers better direct resources to districts in need.

This paper is organized as follows. In Section~\ref{Section: NB approximation}, we explore mathematically how differences in hurdle and ZI model structures are implicitly related to assumptions about disease detection. In Section~\ref{section: modeling-zeros} we detail our proposed coupled Markov switching hurdle model and review its ZI counterpart, the recently proposed coupled Markov switching ZI model \cite{RN5}. Section~\ref{sec: inference} details the Bayesian inferential procedure for both models. In Section 4.2 we include a simulation study comparing predictions between the Markov switching ZI and hurdle models under different reporting rates. We aim to investigate at which reporting rates our proposed Markov switching hurdle model performs better than its zero-inflated counterpart.  Section~\ref{sec: analysis} presents the analysis of the chikungunya data and a comparison in terms of prediction between the hurdle and zero-inflated Markov switching models, and also more conventional zero-inflated and hurdle alternatives. The paper concludes with a discussion in Section~\ref{sec: discussion}.

\section{Motivating a comparison between Zero-Inflated and Hurdle Negative Binomial Models}\label{Section: NB approximation}

In epidemiology, both ZI and hurdle models assume that, for every area and/or time period, the disease can be either present or absent and that when the disease is absent no cases will be reported \cite{RN45, harris_climate_2019}. The main difference between hurdle and ZI models is that, when the disease is present, a hurdle model assumes the reported cases come from a zero-truncated distribution while a ZI model assumes the reported cases are generated by a count distribution that can produce zeroes, such as the negative binomial distribution. { To illustrate how these differences relate to assumptions about disease detection, let $z$ be the {\it actual counts} when the disease is present, and $y$ be the {\it reported counts} when the disease is present. When the disease is present the actual counts $z$ must be greater than zero and so it is reasonable to assume $z$ follows a zero-truncated negative binomial model, that is, }
\begin{equation}
    z|\lambda, r\sim ZTNB(\lambda, r),
    \label{eqn: ZTNB}
\end{equation}
where $\lambda$ is the mean value and $r$ is the over-dispersion parameter of the original negative binomial distribution before the truncation. Then, it is reasonable to assume that the distribution of the reported cases $y$,  {\em given} the actual number of cases $z$, 
follows a binomial distribution, i.e.,
\begin{equation}
    y|z, p_0\sim \text{Binomial}(z,p_0),
\end{equation}
where $p_0$ is the probability of reporting any one case when the disease is present. Under these assumptions, it can be shown that the marginal distribution of $y$ (with respect to $z$) is given by  
\begin{equation}
p(y|\lambda, r, p_0) = \left\{
\begin{array}{lcclcc}
\frac{ (\frac{r}{r+\lambda})^rp_0^y(r+y-1)!(\frac{\lambda}{\lambda+r})^y (\frac{\lambda p_0+r}{\lambda+r})^{-r-y} }{(r-1)! y! (1-(\frac{r}{r+\lambda})^r)} & & &\  y>0,& & \\
\frac{(\frac{r}{\lambda})^r(1-(\frac{\lambda p_0+r}{\lambda+r})^{-r})}{(\frac{r}{\lambda})^r-(\frac{\lambda+r}{\lambda})^r} & & & \   y=0. & & 
\end{array}
\right.
\label{eqn: pmf-of-Z}
\end{equation}
See Section 1 of the Supplementary Material (SM) for the derivation of this distribution. 

Under perfect detection, when $p_0=1$, then $y=z$; that is, $y$ follows a hurdle model with the count part given by equation (\ref{eqn: ZTNB}). If, instead, $p_0<1$, the observed zeroes could
represent the situation wherein the disease is present but is undetected. This leads to a ZI model with count part given by equation (\ref{eqn: pmf-of-Z}).
Although this looks like a reasonable model,  in practice, it cannot be used. This is because the p.m.f. in equation  (\ref{eqn: pmf-of-Z}) involves the reporting probability $p_0$ which is not identifiable from the reported cases alone \cite{RN51}.

Commonly, one chooses a negative binomial random variable  $W$ to accommodate the count part of a ZI model. Therefore,  one can think that the distribution of $W$ approximates the distribution of the {\em true reported counts} $y$\cite{RN55}. To check the appropriateness of this approximation,  one can match  the mean and variance of $W$ with those of the exact distribution in Equation (\ref{eqn: pmf-of-Z}),
\begin{equation}
    W\sim NB(\mu^{(w)}, r^{(w)}),
    \label{eqn: approx-NB}
\end{equation}
where $\mu^{(w)}=\frac{p_0\lambda}{1-(\frac{r}{r+\lambda})^r}$ and $r^{(w)}=\frac{1}{(1-(\frac{r}{r+\lambda})^r)(1+\frac{1}{r})-1}$. See Section 1 of the SM for this derivation. 

\begin{figure}[!t]
    \centering
    \includegraphics[width=16cm]{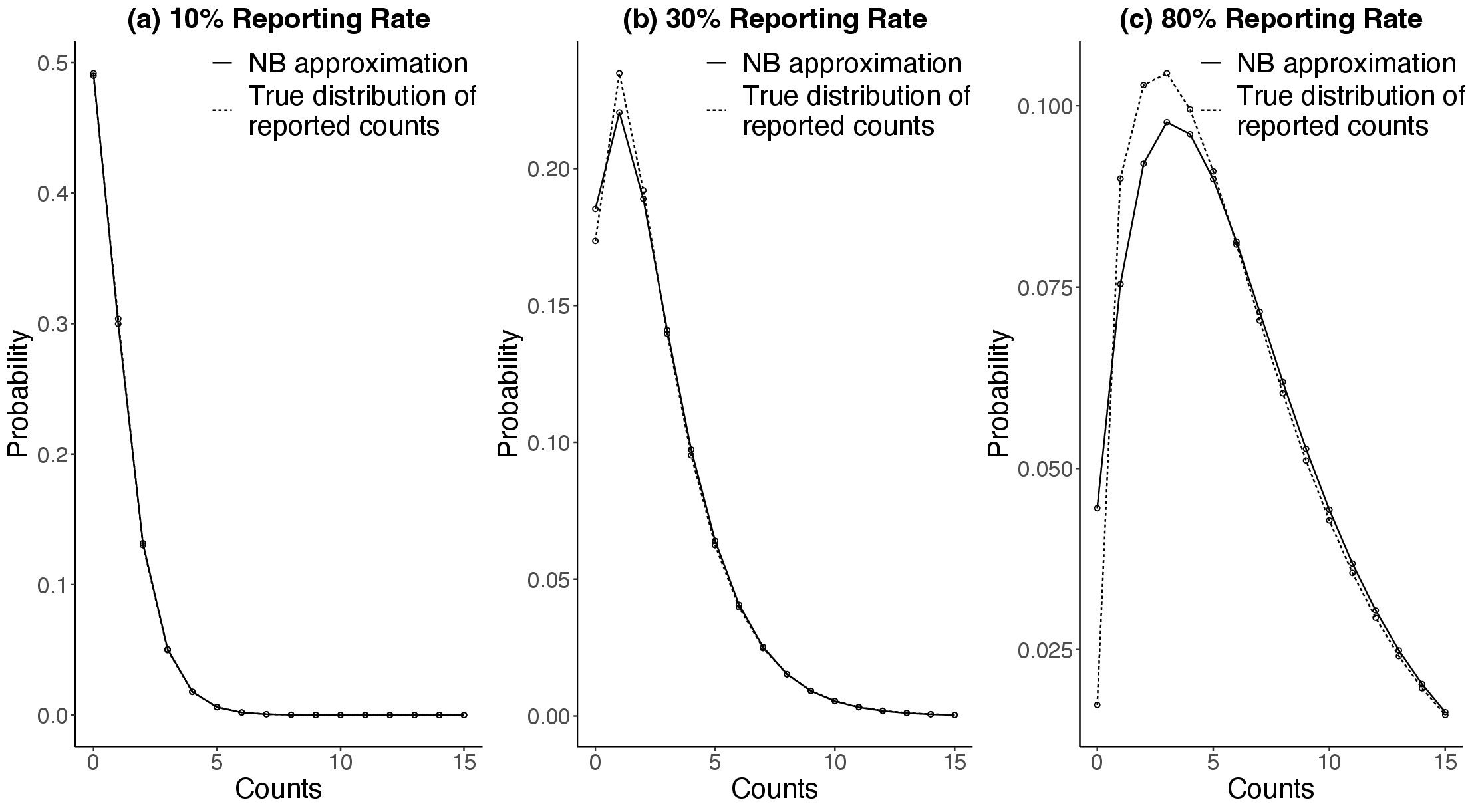}
    \caption{ Probability mass functions 
    of the true distribution of reported counts ($y$) and the approximated NB distribution ($W$) under (a) a low reporting rate ($p_0=0.1$), (b) a moderate reporting rate ($p_0=0.3$) and (c) a large reporting rate ($p_0=0.8$). The over-dispersion of the actual counts ($r=2$) and the incidence of actual cases ($\lambda=8$) are the same for all three scenarios. }
    \label{fig:compare-NB-Z}
\end{figure}

Figure \ref{fig:compare-NB-Z} shows three scenarios of the comparison between the exact distribution in equation (\ref{eqn: pmf-of-Z}) and the approximated negative binomial distribution (\ref{eqn: approx-NB}) of the reported counts. When the reporting rate is small ($p_0=10\%$), using a negative binomial distribution is close to the exact distribution of the reported counts. This suggests that a ZI model with the commonly used negative binomial count part is a reasonable choice under a small reporting rate. However, the approximation becomes poor as the reporting rate increases. This suggests that a hurdle model may be more applicable when the reporting rate $p_0$ is close to $1$. 
Note that this result is intuitive since if the reporting rate is close to 1, we would not expect many zeroes due to a failure to detect the disease even if the expected number of actual cases were small. The negative binomial distribution places a lot of weight on 0 when the mean is small, and thus it would not fit the true reported cases well.

Note that the discussion above is simply to motivate the comparison between ZI and hurdle models when modelling reported cases of a disease. The proposed approaches in the following subsections do not accommodate under-reporting. This is outside the scope of this paper. See Stoner {\em et al.}\cite{RN51} and Oliveira {\em et al.}\cite{lopes2022bias} for examples of models that handle under-reporting. For the zero-inflated models considered here, we only focus on modelling the mean and overdispersion of the reported, not actual, counts, that is, $\mu^{(w)}$ and $r^{(w)}$ in Equation (\ref{eqn: approx-NB}). Since we only model the mean of the reported counts, we cannot tell if a covariate effect is due to changes in disease transmission or reporting rates \cite{douwesschultz2023threestate}. Note that, from Equation (\ref{eqn: approx-NB}), both $\mu^{(w)}$ and $r^{(w)}$ depend on $\lambda$, the mean of the actual counts, which likely varies across space and time. Therefore, in Section~\ref{section: modeling-zeros} below, we assume that both the mean and overdispersion of the Markov switching zero-inflated and hurdle models can depend on covariates.

\section{Modeling Zeros: Zero-inflated and Hurdle Models}\label{section: modeling-zeros}

Assume we have infectious disease cases in areas $i=1,2,..., N$ and at times $t=1,2,..., T$. Let $y_{it}$ be the reported disease case counts from area $i$ at time $t$. Let $X_{it}$ be a binary random variable that indicates the true presence or absence of the disease in area $i$ at time $t$, i.e., $X_{it}=1$ if the disease is present and $X_{it}=0$ if the disease is absent. 
Let $\bm{y}_k=(y_{1k},y_{2k},\cdots, y_{Nk})^T$ be the vector containing the observed counts of the disease across the areas at time $k$. Additionally, let $\bm{y}^{(t-1)}=(\bm{y}_1,\bm{y}_2,\cdots,\bm{y}_{t-1})^T$ be the vector of counts up to time $t-1$. Finally, all available observations are stacked onto the vector
$\bm{y}=(\bm{y}_1,\bm{y}_2, \cdots, \bm{y}_T)^T$.

Under the assumptions of a zero-inflated or hurdle model, when the disease is present, the reported cases are generated by a count distribution, $p(y_{it} \mid \bm{\theta}_{it},X_{it}=1,\bm{y}^{(t-1)})$, and when the disease is absent, no cases are reported. Note that $\boldsymbol{\theta}_{it}$ is the parameter vector defining the distribution of $y_{it}$ when the disease is present. To save space, we will suppress the conditioning on $X_{it}=1$ and $\bm{y}^{(t-1)}$ in $p(y_{it} \mid \bm{\theta}_{it},X_{it}=1,\bm{y}^{(t-1)})$ and denote it as $p(y_{it} \mid \bm{\theta}_{it})$.  Generally, the reported cases for area $i$ at time $t$ given the disease's presence/absence status and the vector of counts up to time $t$ can be expressed as 
\begin{equation}
y_{it}|X_{it},\bm{y}^{(t-1)}\sim \left\{
\begin{array}{lcc}
0 & \ \text{if}\ X_{it}=0 &  (\text{absence}),\\
p(y_{it} \mid \boldsymbol{\theta}_{it}) &  \ \text{if} \ X_{it}=1& (\text{presence}).
\end{array} \label{eqn: counts-structure}
\right.
\end{equation}
As discussed in Section \ref{Section: NB approximation}, a negative binomial distribution is a reasonable specification for $p(y_{it} \mid \boldsymbol{\theta}_{it})$ if the reporting rate is small. Then a zero that comes from the negative binomial distribution represents the disease going undetected, while a zero from the zero mass distribution represents the true absence of the disease. Such models are known as zero-inflated models \cite{RN45}. In contrast to zero-inflated models, if we assume perfect detection of the disease, we can specify $p(y_{it} \mid \boldsymbol{\theta}_{it})$ as a zero-truncated negative binomial distribution since, under perfect detection, there will always be cases reported when the disease is present. Such models are known as hurdle models \cite{harris_climate_2019}.

The practical difference between ZI and hurdle models is that for ZI models, the zeros can come from both disease absence or undetected cases, while in hurdle models zero counts can only be generated due to the actual absence of the disease. That is, hurdle models assume perfect detection of the cases or at least that undetected cases are too few to be relevant. However, a zero-inflated model allows for the imperfect detection of disease cases.  

\subsection{Modeling the Presence/Absence of Disease}

If the disease is present or absent, we would expect it to be more likely to be present or absent again in the next reported time. Therefore, we assume that $X_{it}$, conditional on all the previous cases before time $t$, $\bm{y}^{(t-1)}$, and the presence/absence of the disease in all neighboring areas in the previous time, denoted by $\bm{X}_{(-i)(t-1)}$, follows a two-state non-homogeneous Markov chain. The transition probability matrix of the Markov chain is given by, 
\begin{align}
&\Gamma\left(X_{it}|\bm{y}^{(t-1)},\bm{X}_{(-i)(t-1)}\right) =
\kbordermatrix{
\textbf{State}  & X_{it}=0\ (\textbf{absence}) \,  & &  X_{it}=1\ (\textbf{presence}) \,  \\[8pt]
  X_{i,t-1}=0\ (\textbf{absence}) \, & 1-p01_{it}& & p01_{it} \\[5pt]
  X_{i,t-1}=1\ (\textbf{presence}) \, & 1-p11_{it} & & p11_{it}
  },  \label{eqn:ZSMSP2}
\end{align}
where 
$$
p01_{it}= P(X_{it}=1|X_{i,t-1}=0,\bm{y}^{(t-1)},\bm{X}_{(-i)(t-1)}) \quad   \text{(probability of reemergence)},
$$
$$
p11_{it}= P(X_{it}=1|X_{i,t-1}=1,\bm{y}^{(t-1)},\bm{X}_{(-i)(t-1)}) \quad  \text{(probability of persistence)}.
$$
We also want a statistical model that can investigate how disease persistence and reemergence may be explained by multiple risk factors. Due to the characteristics of infectious diseases, the disease is more likely to be present in an area when the disease is present in its neighboring areas in the previous week. Therefore, the probability of reemergence in area $i$ at time $t$, i.e. $p01_{it}$, can depend on a vector of risk factors $\bm{g}_{it}$ and the spatial neighbors as 
\begin{equation}
\text{logit}(p01_{it})=\alpha_0^{(0)}+\bm{g}_{it}^T\bm{\alpha}^{(0)}+\gamma_1 \sum_{j \in \text{Nei}(i)} X_{j, t-1} \label{eqn:ZSMSP3},
\end{equation}
where $\text{Nei}(i)$ represents the set of all neighboring areas of area $i$, and $\bm{g}_{it}=(g_{it}^{(1)}, g_{it}^{(2)}, \cdots, g_{it}^{(D)})^T$ is a D-dimensional covariate vector. Similarly, the probability of persistence in area $i$ at time $t$, i.e.  $p11_{it}$, is modelled as
\begin{equation}
\text{logit}(p11_{it})=\alpha_0^{(1)}+\delta^{(1)}   \log(y_{i,t-1}+1)+\bm{g}_{it}^T\bm{\alpha}^{(1)}+ \gamma_2 \sum_{j \in \text{Nei}(i)} X_{j, t-1} 
\label{eqn:ZSMSP4},
\end{equation}
where $\log(y_{i,t-1}+1)$ is a term representing the reported case counts for area $i$ at time $t-1$. 
The term $\log(y_{i,t-1}+1)$ is included in the model for $p11_{it}$ because we find it reasonable to assume that the disease will be less likely to go extinct if there are many cases previously. In (\ref{eqn:ZSMSP3}) and (\ref{eqn:ZSMSP4}), $\bm{\alpha}^{(0)}$ and $\bm{\alpha}^{(1)}$ represent the effects of the covariates on the disease's reemergence and persistence probabilities respectively, and they can be different. Note that this is distinct from more classical zero-inflated and hurdle models where $\bm{\alpha}^{(0)}=\bm{\alpha}^{(1)}=\bm{\alpha}$ and $\alpha_0^{(0)}=\alpha_0^{(1)}=\alpha_0$, i.e.,each covariate must have the same effect on the reemergence and persistence of the disease\cite{RN56}. For the Markov chain, we also need to set initial state distributions for the first time in each area, which we denote by $p_0(X_{i0})$  for $i=1,2,..., N$. 

\paragraph{Modelling the parameters of the count part} 
It is assumed that $p(y_{it} \mid \boldsymbol{\theta}_{it})$ in equation (\ref{eqn: counts-structure}) follows either a negative binomial distribution, in the case of the ZI models, or a truncated negative binomial distribution, in the case of the hurdle models, with mean $\mu_{it}$ and overdispersion parameter $r_{it}$ (for the ZTNB  $\mu_{it}$ and $r_{it}$ are the mean and overdispersion parameter of the NB before truncation). 

For infectious disease counts, previous cases are likely to transmit the disease to other individuals, creating new cases. That is, for an area $i$, the previously reported cases, $y_{i,t-1}$, may affect the expected value $\mu_{it}$ of the reported cases $y_{it}$. Thus, we decompose $\mu_{it}$ as in Bauer and Wakefield (2018) \cite{RN29}, that is,
\begin{equation}
    \mu_{it}=\mu_{it}^{AR}y_{i,t-1}+\mu_{it}^{EN}, \label{eqn: mu-structure}
\end{equation} where $\mu_{it}^{AR}$ is the autoregressive rate, which is a multiplier on the previous week's cases that is meant to capture transmission from the previous cases, and $\mu_{it}^{EN}$ is an endemic component meant to capture infectious risk from other sources like the environment and imported cases.  

The autoregressive AR rate $\mu_{it}^{AR}$ is modelled as
\begin{equation}
    \mu_{it}^{AR}=\exp{(b0_{i}+\bm{g}_{it}^T \bm{\beta}^{AR})},
    \label{AR-part}
\end{equation}
where $b0_i|\sigma_{b0}^2\sim_{IID} N(\beta_0^{AR},\sigma_{b0}^2)$ is an area level random effect and $\bm{\beta}^{AR}$ represents the possible effects of risk factors $\bm{g}_{it}$ on $\mu_{it}^{AR}$. The endemic part $\mu_{it}^{EN}$ is modelled as 
\begin{equation}
    \mu_{it}^{EN}=\exp{\left(b_i+\beta_{2}^{EN}sin\left(\frac{t}{52}2\pi\right)+\beta_{3}^{EN}cos\left(\frac{t}{52}2\pi\right)\right)},
    \label{EN-part}
\end{equation} 
where $b_i|\sigma_b^2 \sim_{IID} N(\beta_0^{EN} +\beta_1^{EN}log(N_i),\sigma_{b}^2)$ is an areal level random effect whose mean is a linear function of the population size of the $i$th district. A possible annual seasonal component is modelled by the sine and cosine components \cite{RN34}. It is known that environmental variables such as temperature and precipitation impact the life cycle of the mosquito that transmits chikungunya \cite{RN45}. As we did not have access to these environmental variables in Rio de Janeiro, we include sine/cosine components as a surrogate to account for the seasonal structure that might be present in the data. We expect there to be more reported cases in the summer than in the winter by the strong effects of climate variables on the mosquito's life cycle \cite{RN32}. 

As shown in Equation (\ref{eqn: approx-NB}) the overdispersion parameter of the negative binomial approximation to the true reported counts depends on the expected number of actual counts and so should vary across space and time with covariates. Therefore, we model the overdispersion parameter of the NB and ZTNB distributions, $r_{it}$, as a log-linear function of covariates and past cases,
\begin{equation}
\log(r_{it})=\alpha_0^{(2)}+\bm{g}_{it}^T\bm{\alpha}^{(2)}+\delta^{(2)}log(y_{i,t-1}+1).
    \label{eqn: r-varies}
\end{equation}

In this paper, we will refer to the models defined by the following equations: 

\begin{itemize}
    \item ZINB: Equations (\ref{eqn: counts-structure}) with $p(y_{it}|\bm{\theta}_{it})=NB(\mu_{it},r_{it})$; (\ref{eqn:ZSMSP2})-(\ref{eqn:ZSMSP4}) with $\bm{\alpha}^{(0)}=\bm{\alpha}^{(1)}=\bm{\alpha}$, $\alpha_0^{(0)}=\alpha_0^{(1)}=\alpha_0$ and $\gamma_1=\gamma_2=0$; (\ref{eqn: mu-structure})-(\ref{eqn: r-varies}),

    \item NBH: Equations (\ref{eqn: counts-structure}) with $p(y_{it}|\bm{\theta}_{it})$ following a zero truncated negative binomial distribution, that is, $p(y_{it}|\bm{\theta}_{it})=ZTNB(\mu_{it},r_{it})$; (\ref{eqn:ZSMSP2})-(\ref{eqn:ZSMSP4}) with $\bm{\alpha}^{(0)}=\bm{\alpha}^{(1)}=\bm{\alpha}$, 
    $\alpha_0^{(0)}=\alpha_0^{(1)}=\alpha_0$ and $\gamma_1=\gamma_2=0$; (\ref{eqn: mu-structure})-(\ref{eqn: r-varies}),
    
    \item 
    ZS-MSNB (Zero-state Markov switching negative binomial) 
    \cite{RN5}:  Equations (\ref{eqn: counts-structure}) with $p(y_{it}|\bm{\theta}_{it})=NB(\mu_{it},r_{it})$, (\ref{eqn:ZSMSP2})-(\ref{eqn: r-varies}),

    \item  ZS-MSNBH (Proposed zero-state Markov switching negative binomial hurdle): Equations~(\ref{eqn: counts-structure}) with $p(y_{it}|\bm{\theta}_{it})=ZTNB(\mu_{it},r_{it})$, (\ref{eqn:ZSMSP2})-(\ref{eqn: r-varies}).
\end{itemize}
The ZINB and NBH models represent classical commonly fit versions of ZI and hurdle models \cite{RN1,tawiah_zero-inflated_2021} while the ZS-MSNB and ZS-MSNBH models represent their Markov switching counterparts. The Markov switching models have some important advantages including allowing for separate covariate effects between the reemergence and persistence. They can also more easily account for many consecutive 0s and positive counts since when the disease is in the presence or absence states it is usually more likely to remain there due to the Markov chain \cite{RN5}.

There are some similarities between the specifications of the ZS-MSNB and ZS-MSNBH models. For a specific number of reported cases in district $i$ at time $t$, $y_{it}$, they both assume a latent indicator variable $X_{it}$ to distinguish the case-generating process. However, the indicator variable $X_{it}$ in a ZS-MSNB model is assumed to be not observed when there are zero reported cases\cite{RN49}. In the ZS-MSNB model, both the negative binomial process and the zero process can produce a zero count, which means an observed zero count could be due to either the disease being absent or undetected. These differences in the model specification of $y_{it}|X_{it}$ lead to divergent interpretations. The ZS-MSNBH model assumes perfect detection of the counts, while the ZS-MSNB model allows for the imperfect detection of the disease counts. 

Furthermore, the ZS-MSNB and ZS-MSNBH models are \emph{a priori} plausible for different patterns of case data. When a time series shows switching between long periods of only zero counts and long periods of positive counts, interspersed with some zeros, a ZS-MSNB model is more applicable, like the time series of reported cases shown in Figure \ref{fig: cases}(b). In contrast, for the case where a time series shows switching between long periods of zero counts and long periods of only positive counts, a ZS-MSNBH model may fit better than a ZS-MSNB model. 

\section{Inferential Procedure}\label{sec: inference}

Let $\bm{X}=(\bm{X}_{1},\bm{X}_{2},...,\bm{X}_T)^T$ be the vector of all state indicators, where $\bm{X}_{t}=(X_{1t},X_{2t},...,X_{Nt})^T$. Let  $\bm{\Theta}_0=(\beta_0^{AR}, \bm{\beta}^{AR}, \beta_2^{EN}, \beta_3^{EN}, \beta_0^{EN}, \beta_1^{EN}, \alpha_0^{(0)}, \alpha_0^{(1)}, \alpha_0^{(2)}, \bm{\alpha}^{(0)}, \bm{\alpha}^{(1)}, \bm{\alpha}^{(2)},  \delta^{(1)}, \delta^{(2)}, \gamma_1, \gamma_2, \sigma_b,\sigma_{b0}, \\ \{b_i\}_{i=1}^N,\{b0_{i}\}_{i=1}^N)^T$ be the whole parameter vector apart from state indicators $\bm{X}$. 

In a ZS-MSNBH model, the marginal likelihood function given $\bm{y}$, marginalizing out the state indicators, is given by 

\begin{equation}
    \begin{aligned}
    p(\bm{y}|\bm{\Theta}_0)&= \prod_{i=1}^N\prod_{t=2}^T p(y_{it}|\bm{y}^{(t-1)},\bm{\Theta}_0)\\
    &= \prod_{i=1}^N\prod_{t=2}^T \text{ZTNB}(y_{it}|\mu_{it},r_{it})p01_{it}^{1-I[y_{i,t-1}>0]}p11_{it}^{I[y_{i,t-1}>0]}+\\
    & I[y_{it}=0](1-p01_{it})^{1-I[y_{i,t-1}>0]}(1-p11_{it})^{I[y_{i,t-1}>0]},
    \end{aligned}
\end{equation}
where $I[\bigcdot]$ represents an indicator function; and, $\text{ZTNB}(y_{it}|\mu_{it}, r_{it})$ represents a zero-truncated negative binomial distribution, where the mean and over-dispersion parameters of the associated negative binomial are given by $\mu_{it}$ and $r_{it}$, respectively. We follow the Bayesian paradigm to estimate the parameters of the models. One of the reasons for using the Bayesian approach is because we cannot marginalize out $\bm{X}$ in the ZS-MSNB model, and so it is the only tractable method for that model \cite{RN5}. Further, the Bayesian approach provides uncertainty quantification about the estimates of interest in a straightforward fashion. We assume prior independence among the components of $\bm{\Theta}_0$. Then we specify independent, zero-mean normal prior distributions, with some large variance, for $\bm{\beta}^{AR}$, $\beta_2^{EN}$, $\beta_3^{EN}$, $\beta_0^{EN}$, $\beta_1^{EN}$, $\alpha_0^{(0)}$, $\alpha_0^{(1)}$, $\alpha_0^{(2)}$, $\bm{\alpha}^{(0)}$, $\bm{\alpha}^{(1)}$, $\bm{\alpha}^{(2)}$,  $\delta^{(1)}$, $\delta^{(2)}$, $\gamma_1$ and $\gamma_2$ ; inverse gamma priors for $\sigma^2_{b0}$ and a uniform prior for $\sigma_b$. Regardless of the prior specification, the posterior distribution is not available in closed form. Thus, we will use Markov chain Monte Carlo methods, particularly a Gibbs sampler with some steps of the Metropolis-Hastings algorithm, to draw samples from the resultant posterior distribution. 

In a ZS-MSNB model, the joint likelihood function considering $\bm{X}$ and $\bm{y}$ is given by
\begin{equation}
    \label{eqn:MSZIL}
    p(\bm{y}, \bm{X}|\bm{\Theta}_0)=\prod_{i=1}^N \prod_{t=2}^T p(y_{it}|X_{it}, \bm{y}^{(t-1)}, \bm{\Theta}_{0})\prod_{i=1}^N p(X_{i1})\prod_{t=2}^T p(X_{it}|\bm{X}_{t-1}, \bm{y}^{(t-1)},\bm{\Theta}_0).
\end{equation}
The Gibbs sampler procedure for the ZS-MSNB model is challenging as $\bm{X}$ is not fully observed and $\bm{X}$ cannot be marginalized from the likelihood function. Therefore, we follow a data augmentation algorithm to obtain samples from the posterior distribution of this model \cite{RN5}.

\subsection{Model Comparison Criteria, Temporal Prediction and Missing Values}

In our Bayesian framework, we can use the Watanabe-Akaike information criterion (WAIC) \cite{RN14} to compare different model specifications. For a ZS-MSNBH model, the WAIC is calculated by
\begin{equation}\label{eqn: WAIC}
  \begin{gathered}
   \text{lpdd} = \sum_{i=1}^{N}\sum_{t=2}^T \log\left(\frac{1}{Q-M}\sum_{m=M+1}^Q p(y_{it}|\bm{y}^{(t-1)},{\bm{\Theta}_0}^{[m]})\right), \\
   \text{pwaic} =\sum_{i=1}^{N}\sum_{t=2}^T Var_{m=M+1}^Q \log\left(p(y_{it}|\bm{y}^{(t-1)},{\bm{\Theta}_0}^{[m]})\right),\\
   \text{WAIC} =-2(\text{lpdd}-\text{pwaic}),
  \end{gathered}
\end{equation}
where the superscript $[m]$ denotes a draw from the posterior distribution of the parameter, $M$ is the size of the burn-in period, $Q$ is the size of the MCMC sample, and $Var_{m=M+1}^Q z_m=\frac{1}{Q-M}\sum_{m=M+1}^Q (z_m-\bar{z})^2$ represents the sample variance. The WAIC calculation is different for the ZS-MSNB model. We follow the method where the calculation is conditional on the state in the ZS-MSNB model \cite{RN5}, while the state is marginalized in the ZS-MSBH model as shown in (\ref{eqn: WAIC}). Since it is not easy to integrate out $\bm{X}$ from (\ref{eqn:MSZIL}), applying WAIC to compare a ZS-MSNB model to the other models can be unfair because it has many more parameters \cite{RN42}. Therefore, we only use WAIC for choosing between separate specifications of the same class of models, while we use proper scoring rules, explained in more detail below, for comparing the predictive performance of different models.
A model specification with the lowest WAIC is considered to have the best fit, and two specifications with a difference of 10 or more in WAIC are usually considered to have significant differences. 

Proper scoring rules \cite{RN15} compare different models on the basis of their out-of-sample predictive performance. Scoring rules measure how well the probabilistic forecasts are by assigning scores based on the predictive distribution and the observation \cite{RN15}. One of the most popular proper scoring rules is the ranked probability score (rps) \cite{RN43}. The model with the lowest rps is considered the best predictive model.  

To produce \textup{K}-step-ahead temporal predictions, we used a simulation process, detailed in SM Section~2, to draw multiple samples from the posterior predictive distributions. Let $T_0$ be the final time point that was used for model fitting; the out-of-sample prediction is performed by obtaining a sample from the posterior predictive distribution at time $T_0+k$ for $k=1,2,\cdots, K$, where $K$ is the maximum step we are interested in. A realization from the posterior predictive distribution is denoted as $y_{i,T_0+k}^{[m]}\sim p(y_{i,T_0+k}|\bm{y})$. See SM Section~2 for the Monte Carlo approximations of the posterior predictive distributions for the ZS-MSNB and ZS-MSNBH models.

To compare the models in terms of their ability to predict the cases, we used the ranked probability score approximated by draws from the posterior predictive distributions. The ranked probability score \cite{RN43} for the $\textup{k}$-th step ahead prediction in district $i$ is defined as

\begin{equation}
    \text{rps}( i,T_0, k)=\sum_{j=0}^\infty (P_{i,T_0,k}(j)-{I}[y_{i, \ T_{0}+k}^{(obs)}\leq j ] )^2,
    \label{eqn: rps}
\end{equation}
where $y_{i,T_0+k}^{(obs)}$ is the observed future value for district $i$, and $P_{i,T_0,k}(j)$ is the empirical cumulative distribution function calculated using the draws $y_{i,T_0+k}^{[m]}\sim p(y_{i,T_0+k}|\bm{y})$, evaluated at $j$. The ranked probability score is given by the average ranked probability score over a set of time points from $T_a$ to $T_b$, i.e.,  
\begin{equation}
    \overline{rps}(k)=\frac{1}{N(T_b-T_a+1)}\sum_{i=1}^{N}\sum_{T_0=T_a}^{T_b} \text{rps}(i, T_0, k).
    \label{eqn: averaged-rps}
\end{equation} The model with the lowest $\overline{rps}(k)$ is considered to be the best model at $\textup{k}$-step-ahead prediction for the evaluation period $T_a$ to $T_b$. 

Finally, if observations are missing from the middle part of the time series, we show the proposed ZS-MSNBH can accurately estimate the missing cases. See Section~7 of the SM for details.

\subsection{Simulation Study}\label{sec: simulation}

In Section~\ref{Section: NB approximation}, we showed that a hurdle model implicitly assumes the disease is perfectly detected, while a negative binomial ZI model gives a good approximation to the true distribution of reported counts when reporting rates are low. Therefore, the Markov switching hurdle model should better fit data with high reporting rates and the Markov switching ZI model should better fit data with low reporting rates. Since the reporting rate is not known in a typical application, we designed a simulation study to investigate this hypothesis.

We simplified the models slightly for the simulation study to reduce the computational cost of having to run many simulations. For $i=1, 2, \cdots, 159$ and $t=1, 2, \cdots, 84$, the true cases, $z_{it}$, are simulated from Equation~(\ref{eqn: counts-structure}). We assume $p(\bm{z_{it}}|\theta_{it})=\text{ZTNB}(\mu_{it}, r)$, where $\mu_{it}=\exp(\beta_0+\beta_1 \text{HDI}_{i}+\beta_2 \text{temp}_t)$. Here, $\text{HDI}_{i}$ is the human development index for the ith neighborhood of Rio de Janeiro, see Section~\ref{sec: analysis} below. The covariate $\text{temp}_t$ is the monthly maximum temperature in Rio between 2011 and 2017 \cite{RN5}. We also let the persistence and reemergence probabilities in Equations~(\ref{eqn:ZSMSP3}) and (\ref{eqn:ZSMSP4}) depend on $\text{HDI}_i$ and $\text{temp}_t$. Therefore, we have the coefficients $\bm{\alpha}^{(0)}=(\alpha_{HDI}^{(01)}, \alpha_{\text{temp}}^{(01)})^T$ and $\bm{\alpha}^{(1)}=(\alpha_{HDI}^{(11)}, \alpha_{\text{temp}}^{(11)})^T$. For simplicity, we did not include $\delta^{(1)}$ in the model. The reported cases are then simulated from 
\begin{equation}
    y_{it}|z_{it}\sim \text{Binomial}(p_0,z_{it}), \label{eq:artificialgen}
\end{equation}
where $p_0$ is the reporting rate. 

In the simulation study, we set $\beta_0= 0.5$, $\beta_1=0.1$, $\beta_2=0.4$, $\alpha_0^{(0)}=-3$, 
$\alpha_{\text{HDI}}^{(01)}=1.15$,
$\alpha_{\text{temp}}^{(01)}=1.1$,
$\gamma_1=0.6$,
$\alpha_0^{(1)}=1.5$, $\alpha_{\text{HDI}}^{(11)}=1.18$, $\alpha_{\text{temp}}^{(11)}=1.2$, $\gamma_2=0.3$ and $r=1.5$. This corresponds to the disease being present around $50\%$ of the time.

\begin{figure}[!t]
    \centering
    \includegraphics[width=\linewidth]{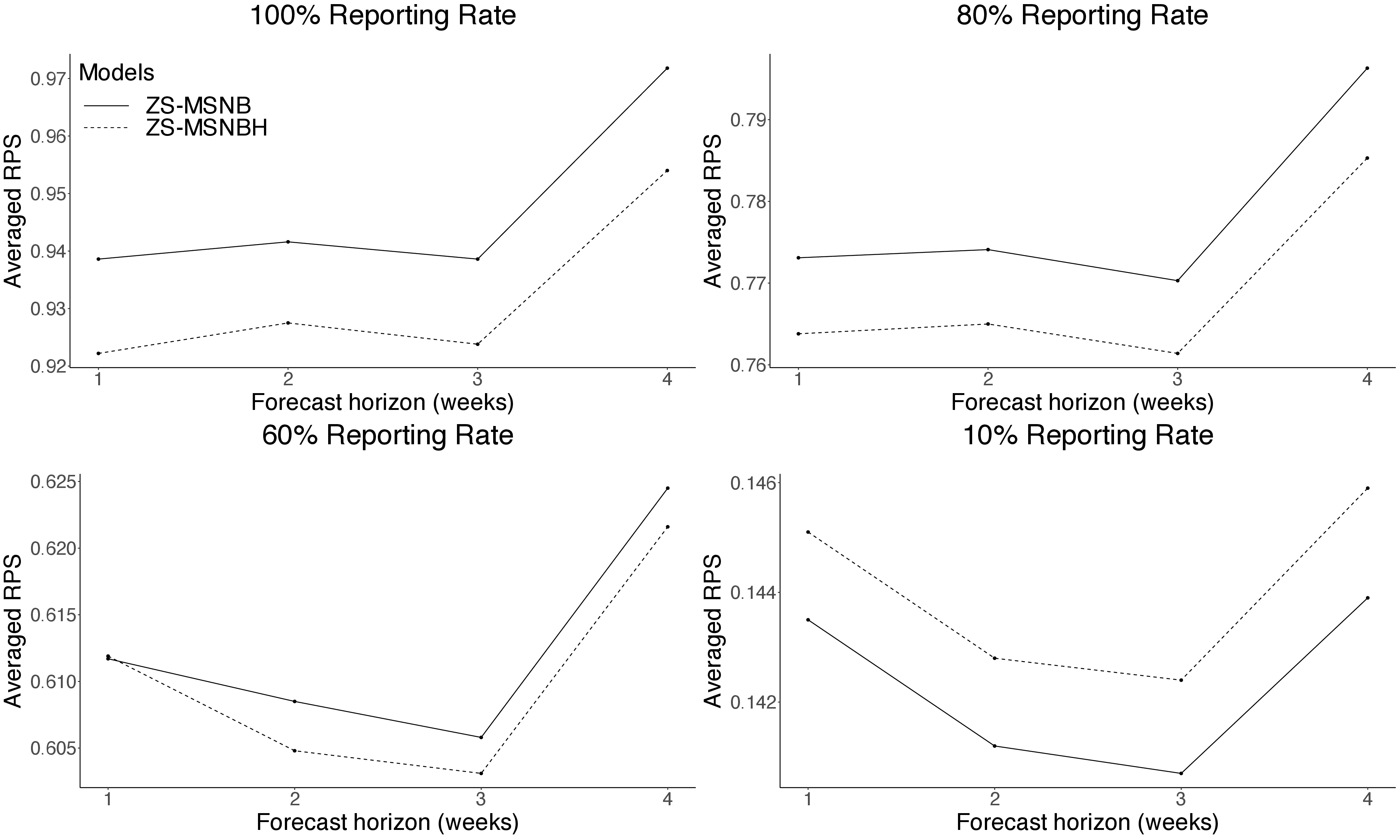}
    \caption{Averaged rps under the 100\% (Top left), 80\% (Top right), 60\% (Bottom left) and 10\% (Bottom right) reporting rates. The solid line represents the ZS-MSNB model, while the dashed line represents the ZS-MSNBH model.} 
    \label{fig:sim-avg-rps}
\end{figure}

Four reporting rates are considered in the simulation study: $p_0^{(1)}=1$, $p_0^{(2)}=0.8$, $p_0^{(3)}=0.6$ and $p_0^{(4)}=0.1$. For each of these scenarios, we simulated data according to the above. Then, we fitted ZS-MSNB and ZS-MSNBH models to investigate which model predicts the reported counts better under each reporting rate, according to proper scoring rules. We used $T_a=40$ to $T_b=80$ as the evaluation period. The ranked probability scores for the Markov-switching ZI and hurdle models are shown in Figures~\ref{fig:sim-avg-rps}, while a comparison of logarithmic scores is shown in SM Section~3 Figure~S1. In Figure~\ref{fig:sim-avg-rps}, the permutation test p-values for the first forecast week under the four reporting rates are 1.55 $\times10^{-5}$ (100\% reporting), 0.002 (80\% reporting), 0.18 (60\% reporting) and 3.8$\times 10^{-6}$ (10\% reporting).

The results from the RPS and logarithmic scores show that the Markov-switching hurdle model gives more accurate predictions, compared to the Markov-switching ZI model, under a large reporting rate (100\%, 80\% reporting). The two models produce similar predictions at a medium reporting rate (60\% reporting) and the Markov switching ZI model produces more accurate predictions under a small reporting rate (10\% reporting). This conclusion agrees with the motivation in Section~\ref{Section: NB approximation}.

\section{Analysis of the Chikungunya Infection Data in Rio de Janeiro}\label{sec: analysis}

In this section, we explore different model structures for the chikungunya dataset, described in Section~\ref{sec:chik}. We first assign a prior distribution to the parameter vector. We assume independent prior distributions; a zero mean normal prior distribution with some large variance, for all unbounded parameters; and we assume independent prior distributions for $\sigma_{b0}^2\sim \text{InvGamma}(0.1,0.1)$ and $\sigma_b\sim \text{Unif}(0,10)$. For the ZS-MSNB model, when $y_{i1}=0$ we assign the prior distribution for the initial state to $p(X_{i1})\sim Bernoulli(0.5)$, while if $y_{i1}>0$ then $X_{i1}=1$. For the ZS-MSNBH model, there is no need to specify the initial distribution of $X_{i1}$ as $X_{i1}=I[y_{i1}>0]$. 

We first investigate the \emph{a priori} plausibility of the ZI/hurdle models based on the model assumptions. When chikungunya was introduced, its circulation was usually not characterized by health authorities, in which case a lot of under-reporting of cases is expected \cite{RN53}. Therefore, the assumptions of the ZINB/ZS-MSNB model, which allows for undetected disease cases, are more plausible than the NBH/ZS-MSNBH model. Also, as discussed in Section 2, the likely low reporting rates suggest that a negative binomial count part for the ZI models is appropriate.

For each of the ZS-MSNB and ZS-MSNBH models, we use WAIC to compare the inclusion/exclusion of the spatial neighbor's terms in Equations (\ref{eqn:ZSMSP3}) and (\ref{eqn:ZSMSP4}), i.e. $\gamma_1$ and $\gamma_2$. As shown in Table~\ref{tab:fitted model-WAIC}, the WAIC supports the inclusion of the spatial terms for the two Markov switching models. Therefore, in this Section, we considered the ZS-MSNB and ZS-MSNBH models, with spatial terms, as well as the NBH and ZINB models, as defined in Section~\ref{section: modeling-zeros}. We also considered a model which assumes the disease is always present, i.e., $X_{it}=1$ for all $i$ and $t$, which we call the negative binomial (NB) model.

\begin{table}[!t]
    \centering
    \caption{Different model specifications when fitted to the chikungunya infectious data compared using WAIC. `Spatial/no spatial' represents the inclusion/exclusion of the spatial neighbors term $\gamma_1$ and $\gamma_2$ (in Equations (\ref{eqn:ZSMSP3}) and (\ref{eqn:ZSMSP4})). The best specification for each model is indicated in italics. }
    \begin{tabular}{llc}
    \toprule
        \textbf{Model} & \textbf{Specification}& \textbf{WAIC}  \\
    \midrule 
        ZS-MSNB & No spatial&   75524.15 \\
      
         & Spatial &  {\em 68450.44}   \\
       ZS-MSNBH & No spatial  &   84958.10 \\
        & Spatial  &  {\em 80379.48}  \\
    \bottomrule
    \end{tabular}
    \label{tab:fitted model-WAIC}
\end{table}

Motivated by our discussion in Section \ref{sec:chik}, the vector of covariates is specified as $\bm{g_{it}}=(\text{HDI}_i,\text{pop}_i, \text{greenarea}_i)^T$, where $\text{HDI}_i$ is the Human Development Index in district $i$, $\text{pop}_i$ is the population in district $i$ obtained from the 2010 Census, the latest available, and $\text{greenarea}_i$ is the proportion of green areas in district $i$ \cite{RN45}. We obtain the Human Development Index data from {\tt ipeadata} ({\tt http://www.ipeadata.gov.br/Default.aspx}), and we obtain the green area data from {\tt datario} ({\tt www.data.rio}). 



The posterior distribution of the fitted models is obtained through MCMC methods as described above and in Section \ref{sec: inference} using the R package \texttt{NIMBLE} \cite{RN52}. For all five models, we ran the Gibbs sampler for 80,000 iterations on 3 chains, with an initial 30,000 iterations considered as burn-in. All the sampling processes began from a random value to avoid local optimization. The codes to run the MCMC are available from GitHub {\tt (https://github.com/MingchiXu/Markov\_Switching\_Hurdle\_code)}. To check the convergence of the chains, we used the Gelman-Rubin statistic (all estimated parameters$<$ 1.05) and the minimum effective sample size ($>$1000) \cite{RN39}. The fitted values in two example districts are shown in Section~4 of the SM for the ZS-MSNB and ZS-MSNBH models. The fitted values were constructed by simulating from the fitted models and show a good agreement between the models and the observed data. However, from SM Section~4, an issue with the ZS-MSNBH model is that it must always switch states when the counts change from positive to zero and vice-versa. This leads to rapid switching between periods of disease presence and absence, which seems unrealistic. Autocorrelation functions (ACF) of the Pearson residuals in two example districts are shown in SM Section~6. No significant autocorrelation is spotted in the ACF plots, which suggests the Markov-switching hurdle model well captured the autocorrelation structure in the data.

Table \ref{tab:posterior summary} shows the posterior summaries from the count part of the ZS-MSNB and ZS-MSNBH models, i.e., equations (\ref{AR-part}) and (\ref{EN-part}) for the two fitted models. The coefficients for the population in both the autoregressive and endemic parts of the mean reported cases are positive, which means higher populated districts have higher transmission of the disease. However, we found there is no evidence of an association between HDI and disease transmission. One possible explanation is that districts with higher HDI are likely associated with higher reporting rates. Therefore, the effects of reporting and disease transmission could cancel each other out. We also did not find an association between green areas and disease transmission. 

\begin{table}[!t]
    \centering
    \caption{Posterior mean and 95\% CI of parameters in the structure of the expected reported cases by different models fitted to the chikungunya data. (A bold-faced estimate means 0 is not included in the 95\% CI) }
    \begin{tabular}{ccc}
    \toprule
        \multirow{2}{*}{ } &
        \multicolumn{2}{c}{\textbf{Posterior mean \& 95\% CI}} \\
        \cmidrule(r){2-3} 
           \textbf{Parameter}  & ZS-MSNB & ZS-MSNBH  \\
          
    \midrule 
       $\makecell[c]{{\beta_0^{AR}}\\ (\text{Intercept})}$
       & \makecell[c]{$-0.379$ \\ (-0.419, -0.341)} & \makecell[c]{$-0.216$ \\ (-0.244, -0.192)}    \\
       
        $\makecell[c]{{\beta_1^{AR}}\\ (\text{pop})}$
        & \makecell[c]{$\bold{0.002}$\\ (0.001, 0.002)} &\makecell[c]{$\bold{0.001}$\\ (0.000, 0.001) }  \\
        
        $\makecell[c]{{\beta_2^{AR}}\\ (\text{HDI})}$
        & \makecell[c]{${0.046}$\\ ({-0.473}, {0.558})} & \makecell[c]{${0.136}$\\ (-0.315, 0.336)}    \\

        $\makecell[c]{{\beta_3^{AR}}\\ (\text{greenarea})}$
        & \makecell[c]{${-0.161}$\\ ({-0.324}, {0.000})} & \makecell[c]{ $-0.038$\\  (-0.149, 0.068)}    \\
        
        $\makecell[c]{{\beta_2^{EN}}\\ (\text{sine})}$
        &  \makecell[c]{$\bold{0.583}$\\ (0.536, 0.630)}  & \makecell[c]{ $\bold{0.65}$\\  (0.566, 0.733) } 
        \\
        
        $\makecell[c]{{\beta_3^{EN}}\\ (\text{cosine})}$
        & \makecell[c]{$\bold{-0.357}$\\ (-0.403, -0.310)} & \makecell[c]{  $\bold{-0.335}$\\ (-0.414, -0.256)}   
        \\
        $\makecell[c]{{\beta_0^{EN}}\\ (\text{Intercept})}$
        & \makecell[c]{$\bold{-0.776}$\\ (-0.860, -0.694)} & \makecell[c]{ $\bold{-1.352}$\\  (-1.465, -1.242) } 
        \\
        $\makecell[c]{{\beta_1^{EN}}\\ (\text{log(pop)} \text{ on Endemic})}$
        &
        \makecell[c]{$\bold{0.435}$\\ (0.357, 0.515)} & \makecell[c]{  $\bold{0.277}$\\  (0.195, 0.363)}  
        \\
   \bottomrule
    \end{tabular}
   \label{tab:posterior summary}
\end{table}

\begin{figure}[!t]
    \centering
    \includegraphics[width=16.5cm]{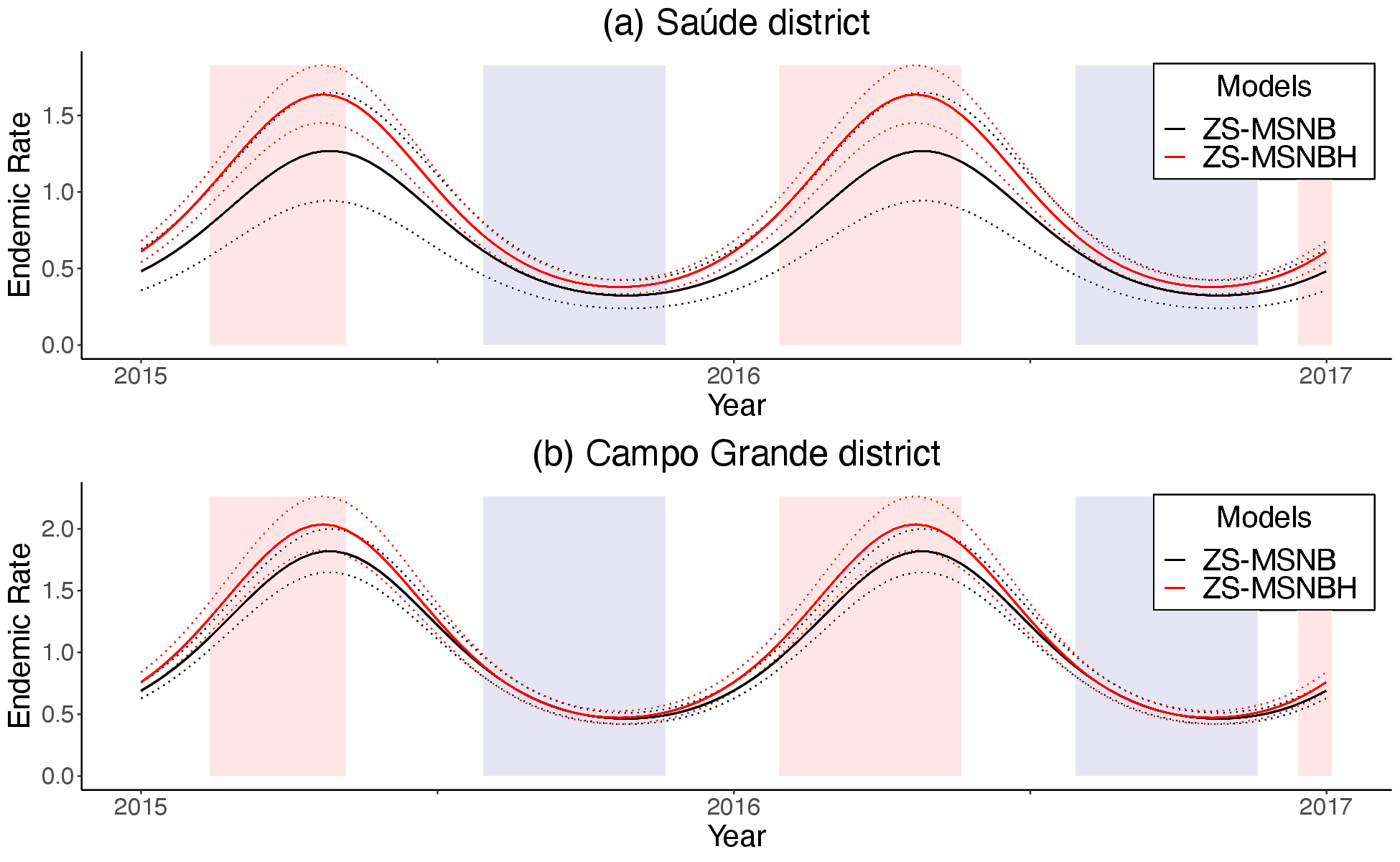}
    \caption{Posterior mean and 95\% credible intervals of endemic rates for (a) Sa\'ude district and (b) Campo Grande district by the ZS-MSNB and ZS-MSNBH models. Summer (December to March)/winter (June to September) seasons are highlighted in red/blue.}
    \label{fig:endemic-parts}
\end{figure}

Figure~\ref{fig:endemic-parts} shows the estimated seasonal trend of the endemic rate under both Markov switching models. The seasonal trend is highest in the summer when mosquito activity is at its highest, and lowest during the winter. The ZS-MSNBH model shows a similar seasonal variation compared to the ZS-MSNB model.

Table~\ref{tab:odds-ratio} presents the odds ratios for the Markov chain part (see equations (\ref{eqn:ZSMSP3}) and (\ref{eqn:ZSMSP4})) of the fitted ZS-MSNB and ZS-MSNBH models. The intercept row can be interpreted as the probabilities of reemergence or persistence assuming one case reported previously for the persistence, keeping other covariates at their average values, and assuming no disease presence in any neighboring areas previously. From Table \ref{tab:odds-ratio}, the estimated probabilities of persistence are much higher than reemergence for the ZS-MSNB and ZS-MSNBH models, meaning the models expect that if the disease is present/absent it will be more likely to be present/absent in the following times. Also in the ZS-MSNBH model, the average probability of staying in the current state (presence or absence) is lower compared to the ZS-MSNB model. This is reasonable because the ZS-MSNBH model is forced to switch states whenever going from positive case counts to zero case counts and vice versa. 

\begin{table}[!t]
    \centering
    \caption{Odds ratio or probability and  95\% posterior credible intervals of parameters in the probabilities of reemergence and persistence for the Markov switching models (including spatial terms) fitted to the chikungunya data. The intercept row shows the probabilities of reemergence or persistence assuming one case reported previously for the persistence, keeping other covariates at their average values, and assuming no disease presence in any neighboring areas previously. A bold-faced estimate means 1 is not included in the 95\% CI.}
    \begin{tabular}{ccccc}
    \toprule
        \multirow{4}{*}{} &
        \multicolumn{4}{c}{\textbf{Probability or Odds ratio}} \\
             &  Persistence  &  & Reemergence &  \\
             &  (presence to presence)  &   & (absence to presence)  & \\
            \cmidrule(r){2-3} \cmidrule(r){4-5} 
             \textbf{Covariates} & ZS-MSNBH & ZS-MSNB  & ZS-MSNBH & ZS-MSNB \\
    \midrule 
         \makecell[c]{\textbf{Intercept}\\ (shifted avg prob)}& \makecell[c]{ $0.226$\\  (0.212, 0.24)}   & \makecell[c]{$0.230$\\ (0.196, 0.267)}   & \makecell[c]{  $0.046$\\ (0.044, 0.048)}  &  \makecell[c]{$0.029$\\ (0.026, 0.032)} 
        \\

         \makecell[c]{\textbf{HDI}\\(.1) }& \makecell[c]{ $1.18$\\  (0.53, 2.272)}  &  \makecell[c]{$0.744$\\ (0.098, 2.706)} &  \makecell[c]{$1.148$\\ (0.671, 1.83)}  &  \makecell[c]{$1.866$\\ (0.553, 4.701)}
        \\
        \makecell[c]{\textbf{Population}\\(1000s) }& \makecell[c]{$\bold{1.004}$\\ (1.003, 1.005)}   &    \makecell[c]{$0.999$\\ (0.997, 1.001)}   & \makecell[c]{ $\bold{1.009}$\\ (1.008 ,1.009)}  &  \makecell[c]{$\bold{1.006}$\\ (1.004, 1.008)}
        \\
       
         \makecell[c]{\textbf{Green area}\\(\%) }& \makecell[c]{$\bold{0.512}$\\ (0.406 , 0.638) } &  \makecell[c]{$1.64$\\ (0.932, 2.759)} &  \makecell[c]{ $\bold{0.328}$\\ (0.28, 0.383)}  &  \makecell[c]{$\bold{0.377}$\\ (0.258, 0.528)}
        \\ 
        \makecell[c]{\textbf{Spatial effects}\\(1) }& \makecell[c]{ $\bold{1.456}$\\ (1.416, 1.496)} &  \makecell[c]{$\bold{2.009}$\\ (1.893, 2.136)} &  \makecell[c]{$\bold{1.99}$\\ (1.946, 2.035)}  &  \makecell[c]{$\bold{3.485}$\\ (3.22, 3.767)}
        \\ 
        \makecell[c]{ $\bold{log(y_{i(t-1)}+1)}$ \\ }& \makecell[c]{ $\bold{7.7}$\\ (6.79, 8.726)}  &  \makecell[c]{$\bold{3.427}$\\ (2.989, 3.911)} &  \makecell[c]{\\ } &  \makecell[c]{\\ }
        \\
        
   \bottomrule
    \end{tabular}
   \label{tab:odds-ratio}
\end{table}

A higher population size is associated with higher odds of chikungunya persistence and reemergence. This means the disease is less likely to go extinct and reemerges quickly in high-population areas, which follows well-known epidemiological theory \cite{RN27}. Also, population size has a larger effect on the reemergence of the disease compared to the persistence. Areas with more green space generally have lower odds of chikungunya reemergence and persistence. Disease presence in neighboring districts previously is associated with higher odds of both disease reemergence and persistence. The previous number of reported counts has a larger effect on disease persistence for the ZS-MSNBH model compared to the ZS-MSNB model. This is likely because the disease can not persist when there are zero cases previously in the hurdle model, which is possible in the ZI model. Therefore, to be relatively equal in the probability of persistence, the Markov switching hurdle model needs to have a larger slope compared to the Markov switching ZI model. Interestingly, HDI has no significant effect on the risk of chikungunya reemergence or persistence according to either model.  

\begin{figure}[!t]
    \centering
    \includegraphics[width=17cm]{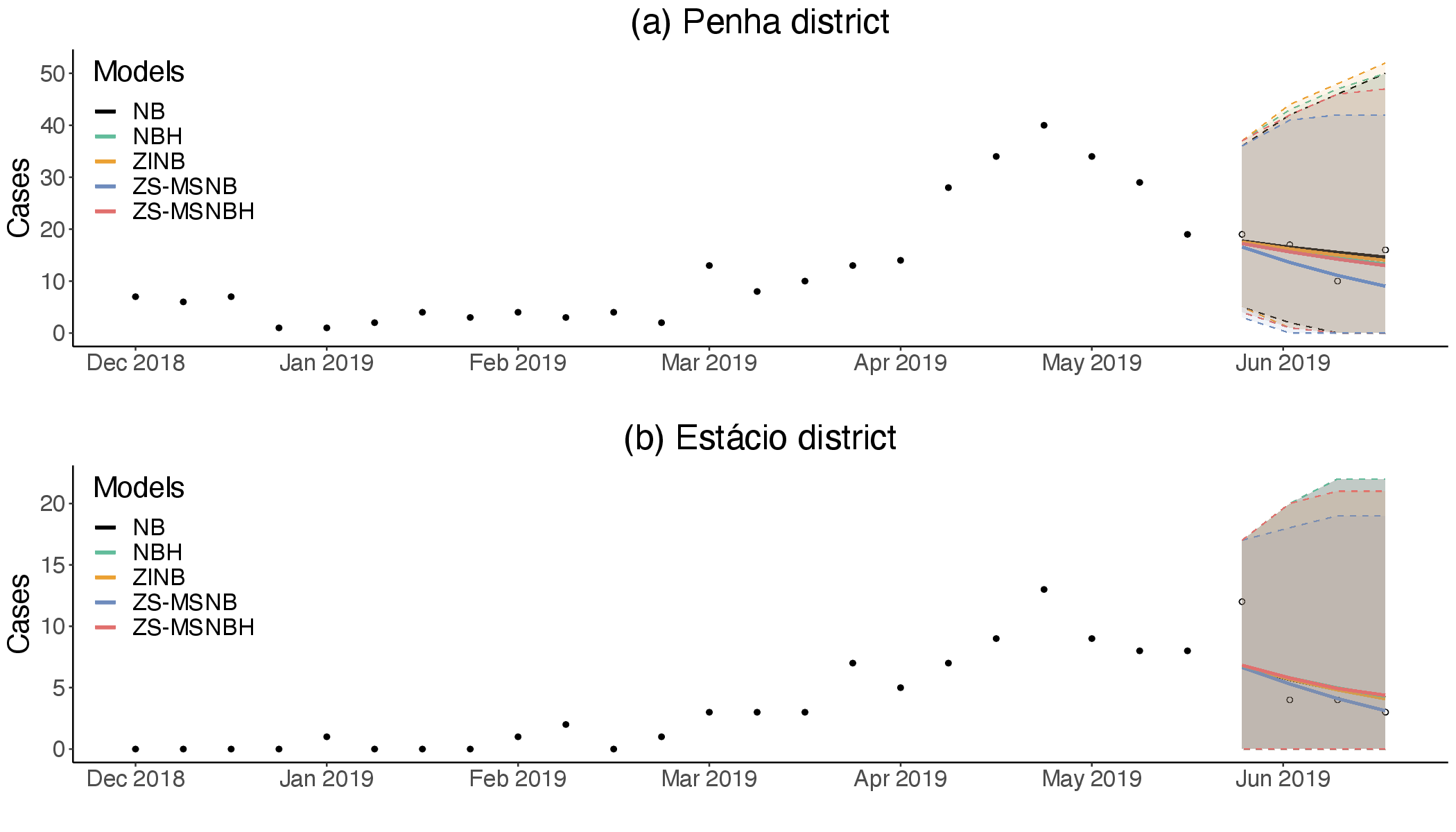}
    \caption{Four steps ahead prediction for (a) Penha district and (b) Estácio district at the observation week $T=231$ by different models and their 95\% prediction interval.}
    \label{fig:mod-pred-comparison}
\end{figure}

Posterior summaries of eight weeks ahead forecasting for the ZS-MSNB and ZS-MSNBH models in two example districts are shown in Section~5 of the SM. Panels of Figure \ref{fig:mod-pred-comparison} show the four steps-ahead predictions at week $T=231$ for the five models considered in this Section (NB, ZINB, NBH, ZS-MSNB, and ZS-MSNBH), for two districts. It seems there are not many differences in the posterior predictive means for the two districts, but different widths of the 95\% credible intervals. The Markov switching ZI model has narrower credible intervals than the other four models, which is likely due to the ZI models switching less between presence and absence compared to hurdle models.

\begin{figure}[!t]
    \centering
   
    \includegraphics[width=16.5cm]{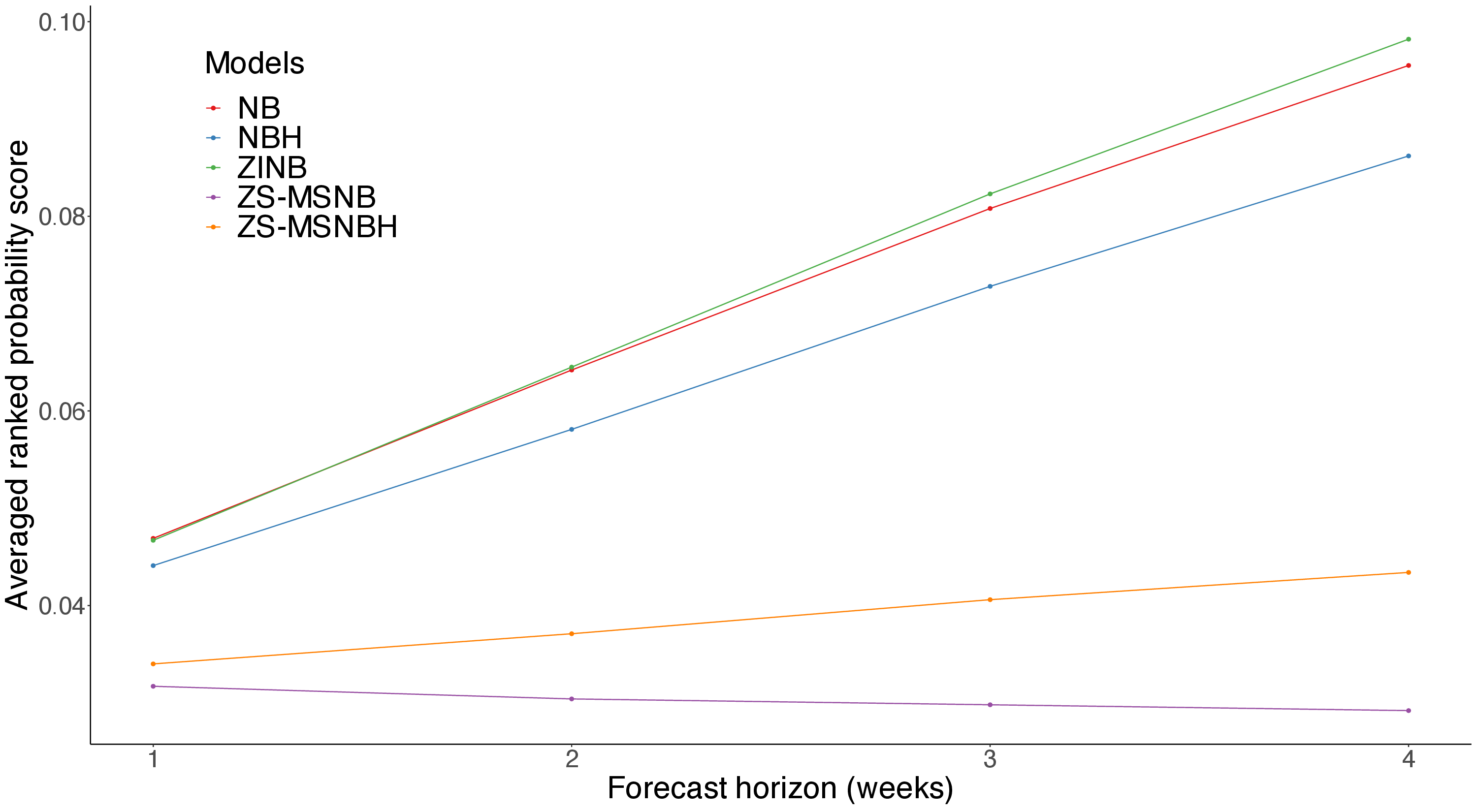}
    \caption{Averaged ranked probability scores across forecast horizons (weeks) according to five models}.
    \label{fig:rps-comparison}
\end{figure}

Figure \ref{fig:rps-comparison} shows the averaged up to four steps-ahead rps for five different models (NB, ZINB, NBH, ZS-MSNB, and ZS-MSNBH). We fit each model up to the time points $T_0=280,281,...,380$, and calculated the 1st to 4th step ahead forecast averaged rps according to (\ref{eqn: rps}) and (\ref{eqn: averaged-rps}). Samples from the models fit up to $T_0$ from $280$ to $380$ all converged according to the Gelman-Rubin statistics \cite{RN39}. The results suggest the NB, NBH, and ZINB models are close to each other in prediction. However, they are all significantly worse than our Markov switching zero-inflated and hurdle models by a large gap at each forecast horizon. Between the two Markov switching models, we performed the permutation test \cite{RN57} and found that the ZS-MSNB model shows a significantly better predictive performance than the ZS-MSNBH model at all forecast horizons (p-values $\textless 2.2\times 10^{-16}$). The test also shows there are significant differences between the best predictive model (the ZS-MSNB model) and the others (all the two-tailed p-values $\textless 2.2\times 10^{-16}$).

\section{Discussion}\label{sec: discussion}

This paper focused on a comparison between Markov switching zero-inflated and hurdle models when accounting for an excess of zeros in spatio-temporal infectious disease counts. The class of zero-state coupled Markov switching negative binomial (ZS-CMSNB) models was recently introduced by Douwes‐Schultz and Schmidt (2022)\cite{RN5}. Their approach assumes the disease switches between periods of presence and absence in each area through a series of coupled Markov chains. This has several advantages over more standard finite mixture zero-inflated approaches such as allowing one to model differently the persistence (presence at time $t-1$ to presence at time $t$) and reemergence (absence at $t-1$ and presence at time $t$) of the disease \cite{RN5}. The ZS-MSNB model can be considered a type of zero-inflated model since it assumes that when the disease is present in an area the reported cases are generated by a negative binomial distribution, which can produce zeroes. 
In contrast, hurdle models are also often used to account for many zeroes in spatio-temporal infectious disease counts and assume differently that the reported cases, when the disease is present, are generated by a zero-truncated distribution \cite{harris_climate_2019}. Therefore, we explored differences in model assumptions and fit between hurdle and zero-inflated models specifically in an epidemiological and Markov switching context. 

In Section 2 we showed that the hurdle model implicitly assumes the disease is perfectly detected while the zero-inflated model assumes there is some non-zero probability of zero reported cases arising due to a failure to detect the disease. Further, we showed that using a negative binomial distribution as the count part of a zero-inflated model, as is common, is only appropriate when the reporting rate is small. Therefore, we would recommend practitioners consider hurdle models for diseases with high detection rates and zero-inflated models for diseases with low detection rates, which is further supported by our simulation study (Section~\ref{sec: simulation}). 

The analyses of the first epidemic of chikungunya experienced in the city of Rio de Janeiro between 2015-2016 was discussed in Section~\ref{sec: analysis}. Different models were fitted to the data, the usual NB, ZINB, and NBH models, together with different structures of the ZS-MSNB and ZS-MSNBH models. Both Markov-switching models produced better predictions compared to the NBH, ZINB, and negative binomial models. Among the fitted models, the ZS-MSNB model was the one with the best predictions according to the ranked probability score. The result is plausible, as the ZS-MSNB model makes the most realistic assumptions by allowing for zeroes due to undetected cases. 
The reporting rates for chikungunya are around 40\% \cite{RN54}, which agrees with our speculation that a ZI model is more plausible for the chikungunya dataset than a hurdle model since, the reporting rate is likely to be small. We also found that chikungunya is more likely to persist and reemerge in areas with high population size, when there is a high amount of neighboring disease presence, when there is less green area, and when there are more cases reported in the previous week. We did not find an important association between HDI and the reemergence or persistence of chikungunya.

There are also some limitations to our approach. 
The Markov switching ZI model assumes the transition probabilities in (\ref{eqn:ZSMSP2}) can depend on covariates and latent disease states in neighboring areas. Therefore, the transition matrix may fluctuate between persistence and non-persistence leading to rapid switching between disease presence and absence, which is not realistic. From examining the fitted values, we found some evidence of rapid state switching in the smaller districts for the Markov switching ZI model, although the problem was much worse for the hurdle model, see Figure~S2. The issue of rapid switching is innate for the Markov switching hurdle model as it always switches states when the counts change from zero to positive or vice-versa. However, for the ZI model, one solution is to introduce "clone states" with determined transitions to enforce minimum disease state durations, like in Douwes-Schultz et al. (2024)\cite{douwesschultz2023threestate}.

\section*{Acknowledgements}

Schmidt is grateful for financial support from the Natural Sciences and Engineering Research Council (NSERC) of Canada (Discovery Grant RGPIN-2017-04999) and Institut de Valorisation des Données (IVADO) (Schmidt -
PRF-2019-6839748021 and Douwes-Schultz PhD-2021-9070375349).

\bibliography{ms.bib}

\end{document}


\maketitle

\startcontents[sections]
\printcontents[sections]{l}{1}{\setcounter{tocdepth}{2}}

\setcounter{equation}{0}
\setcounter{figure}{0}
\setcounter{table}{0}
\setcounter{page}{1}
\makeatletter
\renewcommand{\theequation}{S\arabic{equation}}
\renewcommand{\thefigure}{S\arabic{figure}}
\renewcommand{\bibnumfmt}[1]{[S#1]}
\renewcommand{\citenumfont}[1]{S#1}
\setcounter{section}{0}

\section{Distribution of Exact Probability Mass Function for the Reported Counts}

We provide the calculation for the distribution of the exact probability mass function for $y$ in this part. The actual counts when the disease is present are denoted by $z$, which we assume follows a zero-truncated negative binomial distribution, i.e., 
\begin{equation}
    z|\lambda, r \sim ZTNB(\lambda, r). 
\end{equation}
We assume that the reported counts $y$ follow a binomial distribution conditional on $z$, 
\begin{equation}
    y|z, p_0\sim BIN(z, p_0).
\end{equation}

Firstly, we derive the exact marginal (with respect to $z$) probability mass function for $y$. We consider two cases when $y>0$ and $y=0$. If $y>0$, we have 
\begin{align*}
    p(y)&= \sum_{z=1}^\infty p(y|z)p(z),\\
    &= \sum_{z=1}^\infty \frac{z!}{y! (z-y)!} p_0^y(1-p_0)^{z-y}\times {z+r-1\choose z} \frac{(\frac{r}{r+\lambda})^r(\frac{\lambda}{r+\lambda})^z}{1-(\frac{r}{r+\lambda})^r},\\
    &= \sum_{z=1}^\infty \frac{z!}{y! (z-y)!} p_0^y(1-p_0)^{z-y}\times \frac{(z+r-1)!}{z!(r-1)!} \frac{(\frac{r}{r+\lambda})^r(\frac{\lambda}{r+\lambda})^z}{1-(\frac{r}{r+\lambda})^r},\\
    &= \frac{(\frac{r}{\lambda+r})^r p_0^y (r+y-1)! (\frac{\lambda}{\lambda+r})^y (\frac{\lambda p_0+r}{\lambda+r})^{-r-y}}{(r-1)!y! (1-\frac{r}{\lambda+r})^r}.
\end{align*}

If $y=0$, we have 
\begin{align*}
    p(y)&= \sum_{z=1}^\infty P(y=0|z)p(z),\\
    &= \sum_{z=1}^\infty (1-p_0)^z \times \frac{(z+r-1)!}{z!(r-1)!}\frac{(\frac{r}{r+\lambda})^r (\frac{\lambda}{\lambda+r})^z}{1-(\frac{r}{r+\lambda})^r},\\
    &= \frac{(\frac{r}{\lambda})^r(1-(\frac{\lambda p_0+r}{\lambda+r})^{-r})}{(\frac{r}{\lambda})^r-(\frac{\lambda+r}{\lambda})^r}.
\end{align*}

Secondly, we derive the form for a negative binomial distributed variable, denoted by $W$, used to approximate the exact reported counts distribution, 
\begin{equation}
    W\sim NB(\mu^{(w)}, r^{(w)}).
\end{equation}

To derive a negative binomial distributed variable $W$ with the same mean and variance as $y$, the mean for $W$ should be the same as $y$, i.e.,  
\begin{align*}
    \mu^{(w)}&=E(y) \\
    &= E[E(y|z)]\\
    &= p_0 \times\frac{\lambda}{1-(\frac{r}{r+\lambda})^r}.
\end{align*}
    
Similarly the variance for $W$ should also be the same as $y$, 
\begin{align*}
    Var(W)&=\mu^{(w)}(1+\frac{\mu^{(w)}}{r^{(w)}}) \\
    &= Var(y),\\
    &= E(Var(y|z))+Var(E(y|z)),\\
    &= p_0(1-p_0)E(z)+p_0^2 Var(z),\\
    &= p_0(1-p_0)\frac{\lambda}{1-(\frac{r}{r+\lambda})^r}+ \frac{p_0^2\lambda}{1-(\frac{r}{\lambda+r})^r}\times \frac{1-(1+\frac{r\lambda}{\lambda+r})(1-\frac{\lambda}{\lambda/[1-(\frac{r}{r+\lambda})^r]})}{(\frac{r}{\lambda+r})(\frac{\lambda}{\lambda/[1-(\frac{r}{r+\lambda})^r]})}.
\end{align*}
Therefore we can derive $r^{(w)}$ as follows, 
\begin{align*}
    r^{(w)}&= \frac{(\mu^{(w)})^2}{p_0(1-p_0)\frac{\lambda}{1-(\frac{r}{r+\lambda})^r}+ \frac{p_0^2\lambda}{1-(\frac{r}{\lambda+r})^r}\times \frac{1-(1+\frac{r\lambda}{\lambda+r})(1-\frac{\lambda}{\lambda/[1-(\frac{r}{r+\lambda})^r]})}{(\frac{r}{\lambda+r})(\frac{\lambda}{\lambda/[1-(\frac{r}{r+\lambda})^r]})}-\mu^{(w)}}\\
    &= \frac{ [\frac{p_0\lambda}{1-(\frac{r}{r+\lambda})^r}]^2}{ p_0(1-p_0)\frac{\lambda}{1-(\frac{r}{r+\lambda})^r}+ \frac{p_0^2\lambda}{1-(\frac{r}{\lambda+r})^r}\times \frac{1-(1+\frac{r\lambda}{\lambda+r})(1-\frac{\lambda}{\lambda/[1-(\frac{r}{r+\lambda})^r]})}{(\frac{r}{\lambda+r})(\frac{\lambda}{\lambda/[1-(\frac{r}{r+\lambda})^r]})}-\frac{p_0\lambda}{1-(\frac{r}{r+\lambda})^r}}\\
    &= \frac{1}{(1-(\frac{r}{r+\lambda})^r)(1+\frac{1}{r})-1}.
\end{align*}

\section{Algorithms to Sample from Posterior Predictive Distribution}\label{sec:algorithm}

{For the ZS-MSNB model, we are interested in the k-step ahead posterior predictive distribution of both the disease presence indicator $p(X_{i, T0+k}|\bm{y})$ and the case counts $p(y_{i, T0+k}|\bm{y})$ for $k=1, 2, \cdots, K$ and $i=1, 2, \cdots, N$. We have}{
\begin{multline*}
    \begin{aligned}
        p(X_{i, T0+k}|\bm{y})&=\int p(X_{i, T0+k}|\bm{X}_{T0+k-1}, \bm{y}_{T0+k-1}, \bm{\Theta}_0)\\
        &\times p(\bm{y}_{T0+k-1}|\bm{X}_{T0+k-1}, \bm{y}_{T0+k-2}, \bm{\Theta}_0)p(\bm{X}_{T0+k-1}|\bm{X}_{T0+k-2}, \bm{y}_{T0+k-2}, \bm{\Theta}_0)\\
        &\times \cdots \times p(\bm{y}_{T0+1}|\bm{X}_{T0+1}, \bm{y}_{T0}, \bm{\Theta}_0)p(\bm{X}_{T0+1}|\bm{X}_{T0}, \bm{y}_{T0}, \bm{\Theta}_0)\\
        & \times p(\bm{\Theta}_0|\bm{y})d \bm{y_{T0+k-1}}d\bm{X}_{T0+k-1}\cdots d\bm{y}_{T0+1}d\bm{X}_{T0+1}d\bm{\Theta_0},
    \end{aligned}
\end{multline*}
and 
\begin{multline*}
    \begin{aligned}
        p(y_{i, T0+k}|\bm{y})&=\int p(y_{i, T0+k}|X_{i, T0+k}, \bm{y}_{T0+k-1}, \bm{\Theta}_0)p(X_{i, T0+k}|\bm{X}_{T0+k-1}, \bm{y}_{T0+k-1}, \bm{\Theta_0})\\
        &\times p(\bm{y}_{T0+k-1}|\bm{X}_{T0+k-1}, \bm{y}_{T0+k-2}, \bm{\Theta}_0)p(\bm{X}_{T0+k-1}|\bm{X}_{T0+k-2}, \bm{y}_{T0+k-2}, \bm{\Theta}_0)\\
        &\times \cdots \times p(\bm{y}_{T0+1}|\bm{X}_{T0+1}, \bm{y}_{T0}, \bm{\Theta}_0)p(\bm{X}_{T0+1}|\bm{X}_{T0}, \bm{y}_{T0}, \bm{\Theta}_0)\\
        & \times p(\bm{\Theta}_0|\bm{y})d X_{i, T0+k}d\bm{y}_{T0+k-1}d\bm{X}_{T0+k-1}\cdots d\bm{y}_{T0+1}d\bm{X}_{T0+1}d\bm{\Theta_0}.\\
        &= \int [p(y_{i, T0+k}|X_{i, T0+k}=1, \bm{y}_{T0+k-1}, \bm{\Theta}_0)P(X_{i, T0+k}=1|\bm{X}_{T0+k-1}, \bm{y}_{T0+k-1}, \bm{\Theta}_0)\\
        &+I[y_{i, T0+k}=0](1-P(X_{i, T0+k}=1|\bm{X}_{T0+k-1}, \bm{y}_{T0+k-1}, \bm{\Theta}_0))]\\
        &\times p(\bm{y}_{T0+k-1}|\bm{X}_{T0+k-1}, \bm{y}_{T0+k-2}, \bm{\Theta}_0)p(\bm{X}_{T0+k-1}|\bm{X}_{T0+k-2}, \bm{y}_{T0+k-2}, \bm{\Theta}_0)\\
        &\times \cdots \times p(\bm{y}_{T0+1}|\bm{X}_{T0+1}, \bm{y}_{T0}, \bm{\Theta}_0)p(\bm{X}_{T0+1}|\bm{X}_{T0}, \bm{y}_{T0}, \bm{\Theta}_0)\\
        &\times p(\bm{\Theta}_0|\bm{y})d\bm{y}_{T0+k-1}d\bm{X}_{T0+k-1}\cdots d\bm{y}_{T0+1}d\bm{X}_{T0+1}d\bm{\Theta_0}.
    \end{aligned}
\end{multline*}
The above integrals can be approximated through Monte Carlo integration. {Once samples from $\bm{\Theta}_0$ are available, we can use compositional sampling to sample from $p(X_{i, T0+k}|\bm{y})$ and $p(y_{i, T0+k}|\bm{y})$}. Specifically, the Monte Carlo approximations are  
\begin{equation*}
    p(X_{i, T0+k}|\bm{y})\approx \frac{1}{Q-M}\sum_{m=M+1}^{Q} p(X_{i, T0+k}|\bm{X}_{T0+k-1}^{[m]}, y_{i,T0+k-1}^{[m]}, \bm{\Theta}_0^{[m]}),
\end{equation*}
and 
\begin{multline}
    \begin{aligned}
    p(y_{i, T0+k}|\bm{y})&\approx \frac{1}{Q-M} \sum_{m=M+1}^Q [p(y_{i, T0+k}|X_{i, T0+k}=1, y_{i,T0+k-1}^{[m]}, \bm{\Theta}_0^{[m]})\times \\
    & P(X_{i, T0+k}=1|\bm{X}_{T0+k-1}^{[m]}, y_{i,T0+k-1}^{[m]}, \bm{\Theta_0}^{[m]})\\
    &+I[y_{i, T0+k}=0](1-P(X_{i, T0+k}=1|\bm{X}_{T0+k-1}^{[m]}, y_{i,T0+k-1}^{[m]}, \bm{\Theta}_0^{[m]}))],
\end{aligned}
\label{eqn: MCMC-approx-MSZI}
\end{multline}
for $i=1, \cdots, N$, where the superscript $[m]$ denotes a draw from the posterior distribution of the parameter, $M$ is the size of the burn-in sample and $Q$ is the total MCMC sample size.}  

\begin{algorithm}
\caption{Posterior predictive sampling process for $y_{i,T_0+k}^{[m]}$ in the ZS-MSNB model}
For $i=1,2,...,N$; $k=1,2,...,K$; $m=1,2,...,M$:\\

\begin{itemize}
    \item $\bm{Step\ 1}$: Draw $X_{i,T0+k}^{[m]}\sim p(X_{i,T_0+k}|\bm{X}_{T_0+k-1}^{[m]},y_{i,T_0+k-1}^{[m]},\bm{\Theta}_0^{[m]})$, where $\bm{y}_{T_0}^{[m]}=\bm{y}_{T_0}$,
    \item $\bm{Step\ 2}$: Draw $y_{i,T0+k}^{[m]}\sim p(y_{i,T_0+k}|X_{i,T_0+k}^{[m]},y_{i,T_0+k-1}^{[m]},\bm{\Theta}_0^{[m]})$, where $\bm{y}_{T_0}^{[m]}=\bm{y}_{T_0}$.
\end{itemize}
\end{algorithm}

{For the ZS-MSNBH model, we are only interested in the k-step ahead posterior predictive distribution of the case count $p(y_{i, T0+k}|\bm{y})$, as the disease present indicators are known from the case counts. Similarly, for $k=1, 2, \cdots, K$ and $i=1, 2, \cdots, N$, we have}{
\begin{multline*}
    \begin{aligned}
    p(y_{i, T0+k}|\bm{y})
    &= \int p(y_{i, T0+k}|\bm{y}_{T0+k-1}, \bm{\Theta}_0)p(\bm{y}_{T0+k-1}|\bm{y}_{T0+k-2}, \bm{\Theta}_0)\\
    &\times \cdots \times p(\bm{y}_{T0+1}|\bm{y}_{T0}, \bm{\Theta}_0)p(\bm{\Theta}_0|\bm{y})d\bm{y}_{T0+k-1}\cdots d\bm{y}_{T0+1}d\bm{\Theta}_0,
    \end{aligned}
\end{multline*}
and the above integral can be approximated through Monte Carlo integration,
\begin{equation}
    p(y_{i, T0+k}|\bm{y})\approx \frac{1}{Q-M}\sum_{m=M+1}^Q p(y_{i, T0+k}|\bm{y}_{T0+k-1}^{[m]}, \bm{\Theta}_0^{[m]}).
    \label{eqn: MCMC-approx-MShurdle}
\end{equation}
}

\begin{algorithm}
\caption{Posterior predictive sampling process for $y_{i,T_0+k}^{[m]}$ in the ZS-MSNBH model}

For $i=1,2,...,N$; $k=1,2,...,K$; $m=1,2,...,M$: \\
\begin{itemize}
    \item $\bm{Step\ 1}$: Draw $y_{i,T0+k}^{[m]}\sim p(y_{i,T_0+k}|\bm{y}_{T0+k-1}^{[m]},\bm{\Theta}_0^{[m]})$, where $\bm{y}_{T_0}^{[m]}=\bm{y}_{T_0}$.
\end{itemize}
\end{algorithm}
\newpage

\section{{Averaged Logarithmic Scores in the Simulation Study}}\label{sec:SM-logscores}

{In Section~4.2 we compared the predictions between the Markov switching hurdle and Markov switching ZI models under different reporting rates. The posterior predictive distribution for the ZS-MSNBH model is approximated by Equation~(\ref{eqn: MCMC-approx-MShurdle}). In contrast, the ZS-MSNB model approximates it by Equation~(\ref{eqn: MCMC-approx-MSZI}). 
The logarithmic score\cite{RN15} for the $k$-th step ahead prediction in district $i$ is defined as
\begin{equation}
    \text{logscore}(i, T_0, k)=-\log(p(y_{i,T_0+k}^{(obs)}|\bm{y})),
    \label{eqn: def-logscore}
\end{equation}
where $y_{i,T_0+k}^{(obs)}$ is the observed future value for district $i$, and $p(y_{i,T_0+k}^{(obs)}|\bm{y})$ can be approximated by (\ref{eqn: MCMC-approx-MSZI}) or (\ref{eqn: MCMC-approx-MShurdle}) depending on the model being considered and using the draws from the respective posterior predictive distribution. We then average the logarithmic score over a set of time points from $T_a$ to $T_b$ and all areas to compute the mean scores, i.e.,
\begin{equation}
    LS_{k}=\frac{1}{N(T_b-T_a+1)}\sum_{i=1}^{N}\sum_{T_0=T_a}^{T_b} \text{logscore}(i,T_0,k),
\end{equation}
where $T_a$ and $T_b$ denote the upper and lower bounds for this set of time points. The model with the lowest $LS_{k}$ is considered to be the best model at $\textup{k}$-step-ahead prediction for the evaluation period $T_a$ to $T_b$.} 

\begin{figure}[!htbp]
    \centering
    \includegraphics[width=\linewidth]{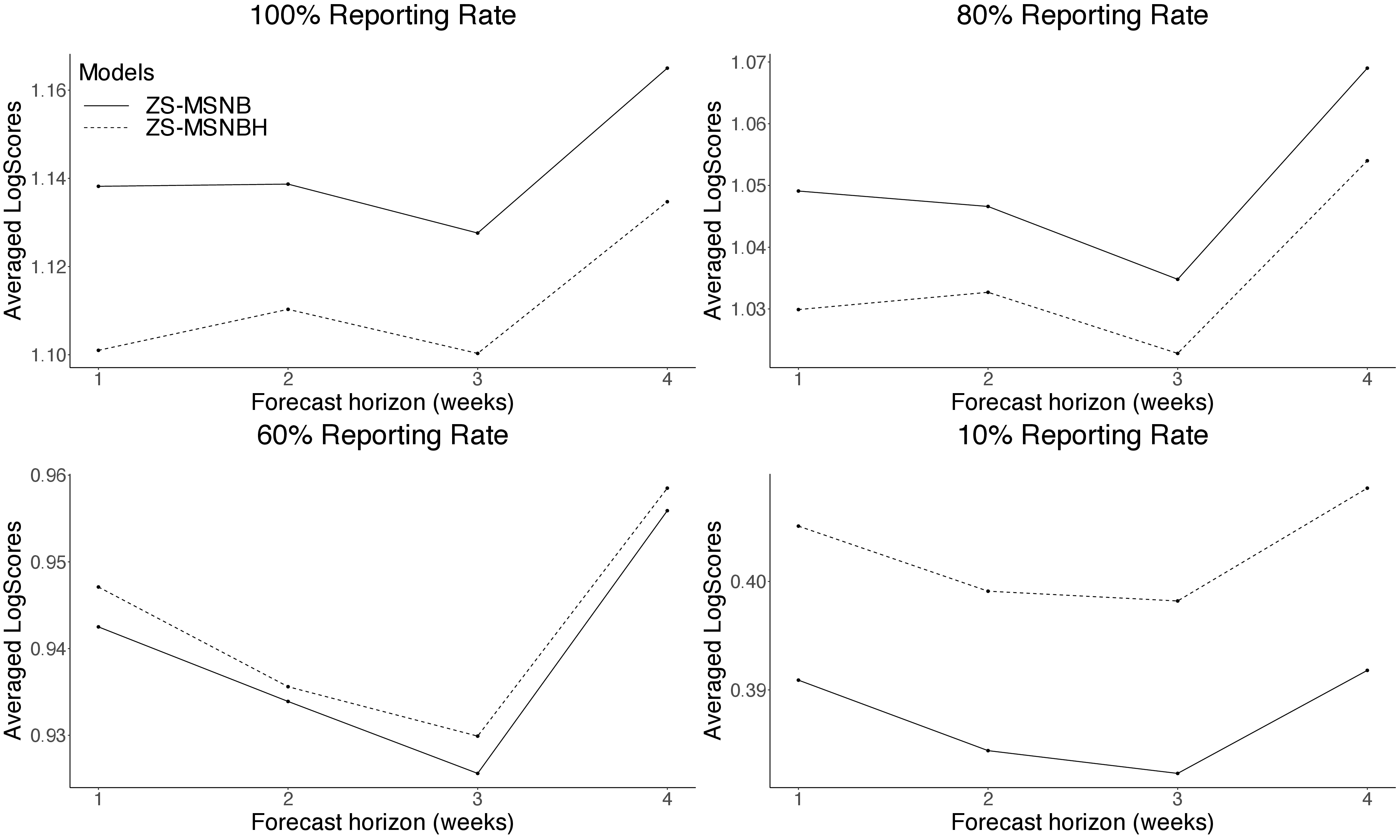}
    \caption{{Averaged log scores under the 100\% (Top left), 80\% (Top right), 60\% (Bottom left) and 10\% (Bottom right) reporting rates. The solid line represents the ZS-MSNB model, while the dashed line represents the ZS-MSNBH model.} }
    \label{fig:sim-avg-logscore}
\end{figure}

{Figure~\ref{fig:sim-avg-logscore} shows the comparison of logarithmic scores between the ZS-MSNB and ZS-MSNBH models, as discussed in Section~4.2. In Figure~\ref{fig:sim-avg-logscore}, the permutation tests p-values for the first forecast week are 2.18 $\times10^{-5}$ (100\% reporting), 0.002 (80\% reporting), 0.88 (60\% reporting) and 0.014 (10\% reporting). The result of the logarithmic scores suggests the same conclusion as that of RPS in Section~4.2. The Markov-switching hurdle model is preferable to the Markov-switching ZI model under a large reporting rate and is unpreferred under a small reporting rate.}

\section{Fitted Values in Two Districts}\label{supp-sec: fitted val}
\begin{figure}[!htbp]
 	\centering
  \includegraphics[width=\textwidth]{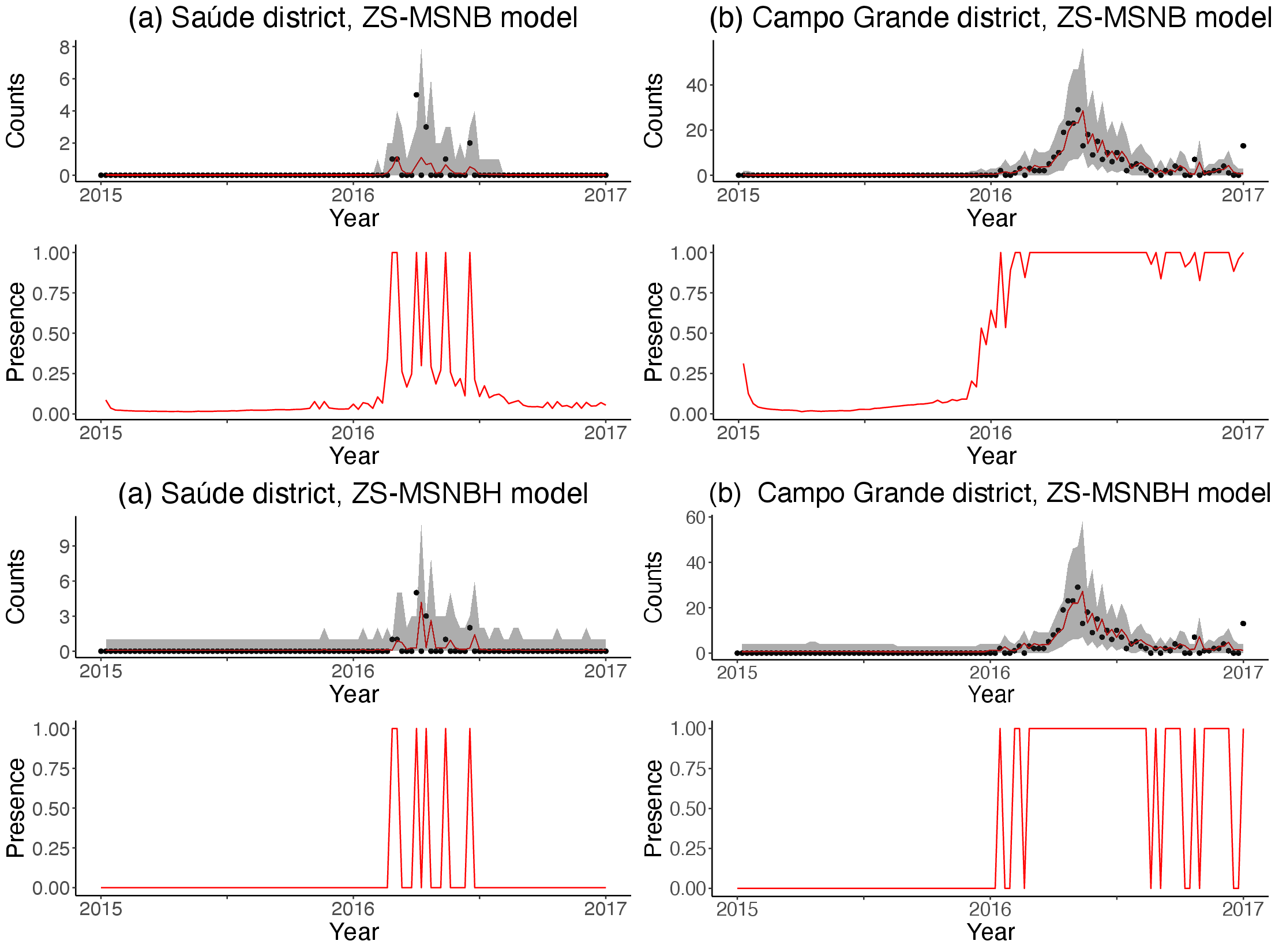}
	\caption{ Posterior summaries of the one-week head fitted values for (a) Sa\'ude district and (b) Campo Grande district by two Markov switching models.(top graphs) One week ahead fitted values of cases versus observed cases. Posterior means (solid lines), 95\% posterior credible intervals (shaded areas) and observed (solid circles). (bottom graphs) Posterior probability of chikungunya presence. } 
 \label{fig: fitted-ex}
\end{figure}

{To investigate if the Markov switching ZI and hurdles model have a good agreement with the data,} we construct and compare the one-week ahead fitted values. Figure \ref{fig: fitted-ex} shows posterior summaries of the one-week ahead fitted values from the two Markov switching models in two neighborhoods. The top graphs show the mean and 95\% quantile of $y_{it}^{[m]} \sim p(y_{it}|\bm{y}^{(t-1)},\bm{\Theta}_0^{[m]})$, for the ZS-MSNBH model, and $y_{it}^{[m]} \sim p(y_{it}|\bm{y}^{(t-1)},\bm{\Theta}_0^{[m]},X_{it}^{[m]})$, for the ZS-MSNB model, from the posterior draws $m=1,2,\cdots, M$. These approximate the posterior mean and credible interval of a new count produced by the model conditional on the observed past counts, parameters and disease state (the latter only in the case of the ZI model) \cite{RN5}. The bottom graphs show the posterior probability of chikungunya presence in district $i$ at week $t$, $P(X_{it}=1|\bm{y})$, which is approximated by the mean of $X_{it}^{[m]}$ from the draws $m=1,2,\cdots,M$. The Markov switching models seem to capture the data well because the actual case counts often fall in the 95\% CI. {However, there is some rapid switching between the disease states, which may not be realistic, as discussed in Sections 5 and 6 of the main text.}

\newpage

\section{Posterior Predictions of the Markov Switching Models in Two Districts}

\begin{figure}[!htbp]
    \centering
    \includegraphics[width=17cm]{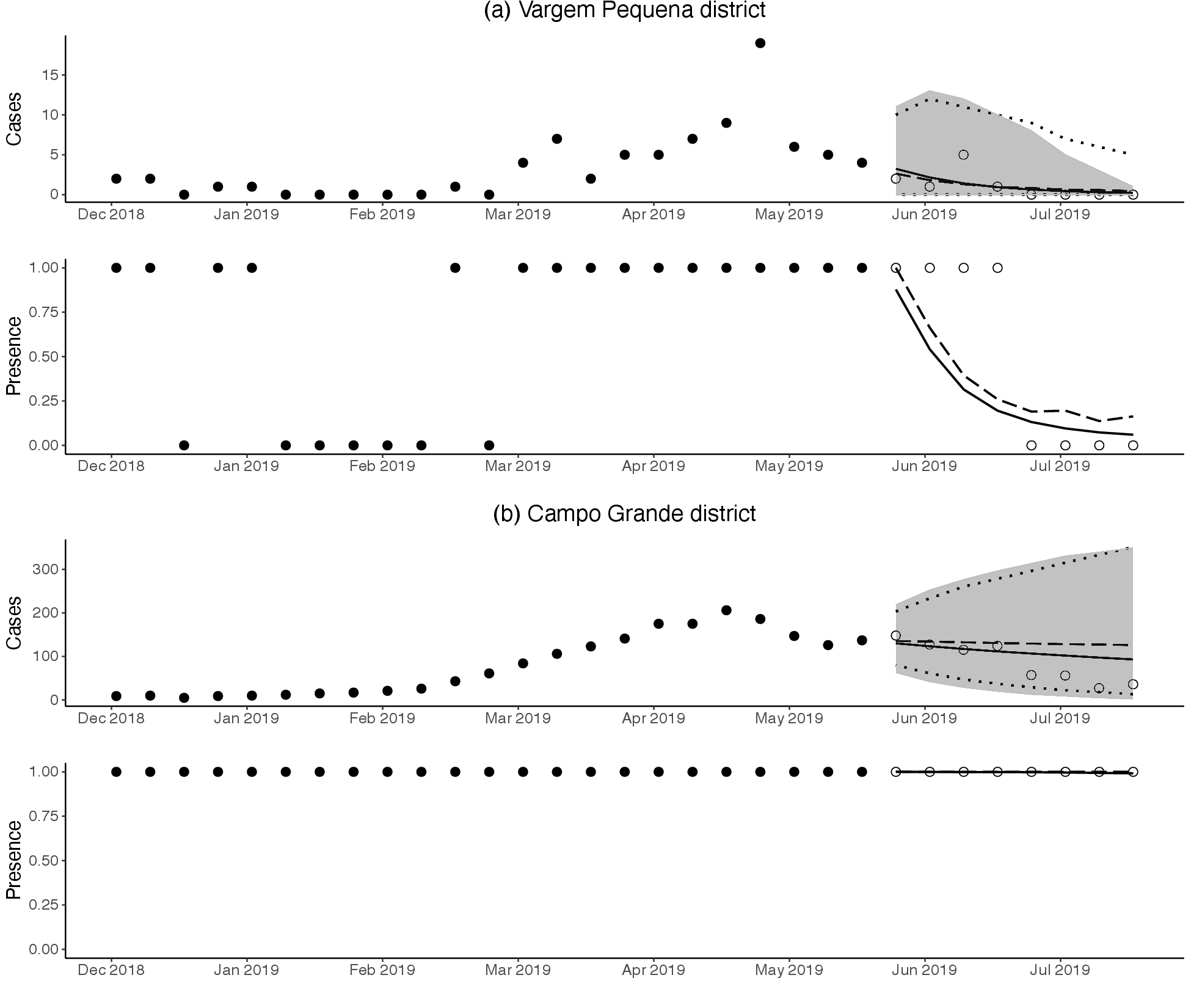}
    \caption{{Posterior summaries of the eight weeks ahead forecasting for the two Markov switching models at the observation week $T=231$ in (a) Vargem Pequena district and (b) Campo Grande district. Solid lines represent the posterior mean and shaded areas represent the 95\% predictive credible interval for the ZS-MSNB model. Dashed lines represent the posterior mean and dotted boundaries represent the 95\% predictive credible interval for the ZS-MSNBH model. Solid circles represent observations of cases (top graphs) and disease presence (bottom graphs) used in the inferential process while open circles represent the eight observations of cases (top graphs) and disease presence (bottom graphs) left out from the inference procedure.}}
    \label{fig:ms-pred-comparison}
\end{figure}

Panels of Figure \ref{fig:ms-pred-comparison} show the posterior summaries of eight weeks ahead forecasting at week $T=231$ for the ZS-MSNB and ZS-MSNBH models, in two districts. Both the two Markov switching models are able to predict the presence of the disease and case counts well, and they can accommodate switching between periods of disease presence and absence.

\section{{Pearson Residuals}}\label{sec:app-pearson-residual}

{One common diagnostic tool in the time series data is the autocorrelation function (ACF) of the Pearson residuals, for example, in Douwes-Schultz and Schmidt (2022)\cite{RN5}. Following Section \ref{supp-sec: fitted val}, we define Pearson's residual for area $i$ as 
\begin{equation}
    \label{eqn:pearson1}
    Res_{it}=\frac{y_{it}-E[y_{it}|\bm{y}^{(t-1)}, \bm{\Theta}_0^{}]}{\sqrt{Var(y_{it}|\bm{y}^{(t-1)}, \bm{\Theta}_0^{})}}
\end{equation}
for the ZS-MSNBH model, and 
\begin{equation}
    \label{eqn:pearson2}
    Res_{it}=\frac{y_{it}-E[y_{it}|\bm{y}^{(t-1)}, \bm{\Theta}_0^{}, X_{it}^{}]}{\sqrt{Var({y}_{it}|\bm{y}^{(t-1)}, \bm{\Theta}_0^{}, X_{it}^{})}}
\end{equation}
for the ZS-MSNB model.}

\begin{figure}[!ht]
    \centering
    \includegraphics[width=11cm]{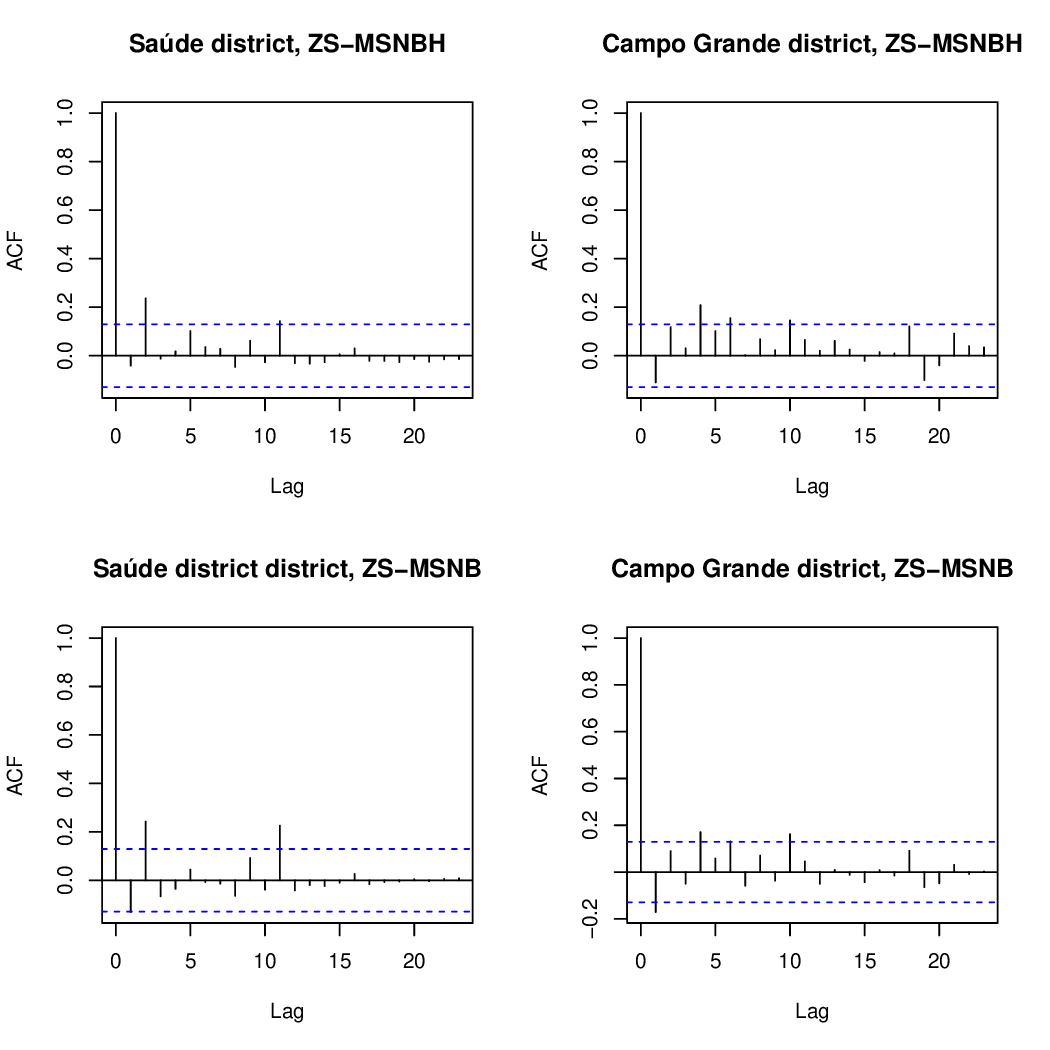}
    \caption{{ACF of the Pearson residuals by two Markov switching models for the 2 districts from Figure S2.}}
    \label{fig:ACF-res}
\end{figure}

{{We estimate the means and variances in Equations (\ref{eqn:pearson1})-(\ref{eqn:pearson2})  using the sample mean and variance of $y_{it}^{[m]}$}. Figure~\ref{fig:ACF-res} shows the ACF of the Pearson residuals for the two districts (Sa\'ude and Campo Grande). Not many significant autocorrelations are present in the ACF plots, which suggests the two Markov switching models capture the correlation structure well.}

\newpage

\section{{Estimating Missing Values}}\label{sec:missing-in-middle}

{Following the suggestion of a reviewer, here we investigate how the models perform when there is missing data. 
From a Bayesian point of view, if observations are missing they become parameters of the model. When observations are missing either completely at random (MCAR) or at random (MAR), commonly, given a sample from the posterior distribution of the parameters, one uses compositional sampling to sample from the distribution that defines the model for each posterior sample, and a posterior sample from the missing value is obtained. In the proposed approach we consider observations to be MAR and the posterior full conditional distribution of the missing value is more complicated than simply sampling from the distribution that is assumed for the data. This is because we have an auto-regressive component in the mean structure of the negative binomial distribution.

In particular, assume that $y_{it}$ is missing. To obtain a sample from the posterior of $y_{it}$, we need to sample from the posterior full conditional distribution of $y_{it}$ at every iteration of the Gibbs sampler. In this case, the posterior full conditional is given by 
\begin{multline*}
    \begin{aligned}
        p(y_{it} \mid \bm{y}_{(-it)}, \Theta) &\propto p(\bm{y}\mid \Theta)\\
        & \propto p(y_{it}|\bm{y}_{t-1}, \Theta)\prod_{j=1}^N p(y_{j (t+1)}|\bm{y}_{t}, \Theta)\\
        &\propto p(y_{it}|\bm{y}_{t-1},\Theta)p(y_{i (t+1)}|\bm{y}_{t}, \Theta)\prod_{j: i\in Nei(j)}p(y_{j (t+1)}|\bm{y}_t, \Theta),
    \end{aligned}
\end{multline*}
where $\bm{y}_t=(y_{1t}, \cdots, y_{Nt})^T$ and $\bm{y}_{(-it)}$ contains all other responses except for $y_{it}$. By the definition of a ZS-MSNBH model, we have
\begin{multline*}
    \begin{aligned}
        p(y_{it}|\bm{y}_{t-1}, \Theta)&=\text{ZTNB}(y_{it}; \mu_{it}, r_{it})p01_{it}^{1-I[y_{i,t-1}>0]}p11_{it}^{I[y_{i,t-1}>0]}\\
        &+I[y_{it}=0](1-p01_{it})^{1-I[y_{i,t-1}>0]}(1-p11_{it})^{I[y_{i,t-1}>0]},\\
    \end{aligned}
\end{multline*}
and we model the persistence and reemergence probabilities $p11_{it}$ and $p01_{it}$ as 
\begin{multline*}
    \begin{aligned}  \text{logit}(p11_{it})&=\alpha_{0}^{(1)}+\delta^{(1)}\log(y_{i,t-1}+1)+\bm{g}_{it}^T \alpha^{(1)}+\gamma_2\sum_{j\in Nei(i)}I[y_{i,t-1}>0],\\
\text{logit}(p01_{it})&=\alpha_{0}^{(0)}+\bm{g}_{it}^T \alpha^{(0)}+\gamma_1\sum_{j\in Nei(i)} I[y_{i,t-1}>0].\\
    \end{aligned}
\end{multline*}
\texttt{Nimble} uses slice samplers to sample from the posterior full conditional distribution $p(y_{it}|\bm{y_{(-it)}}, \Theta)$.


To check the ability of the model to estimate missing values, we reanalyzed the number of cases of chikungunya across the districts of Rio de Janeiro and removed the observations from weeks 57-59, 61-62, and 71-72 from the "Centro" district. Then we fit the Markov switching hurdle model and estimated the missing values. Figure \ref{fig:missing-in-mid} shows the observed time series in the  "Centro" district, together with the values that were removed from the inference (open circles) and the pointwise summaries (mean and limits of the 95\% posterior credible intervals) at the times that were removed from the series. {The figure shows the posterior mean for the missing points is close to the true values, and the 95\% credible intervals mostly include the removed points. Our proposed ZS-MSNBH model can estimate the missing cases well. }
}

{\begin{figure}[!htbp]
    \centering
    \includegraphics[width=\linewidth]{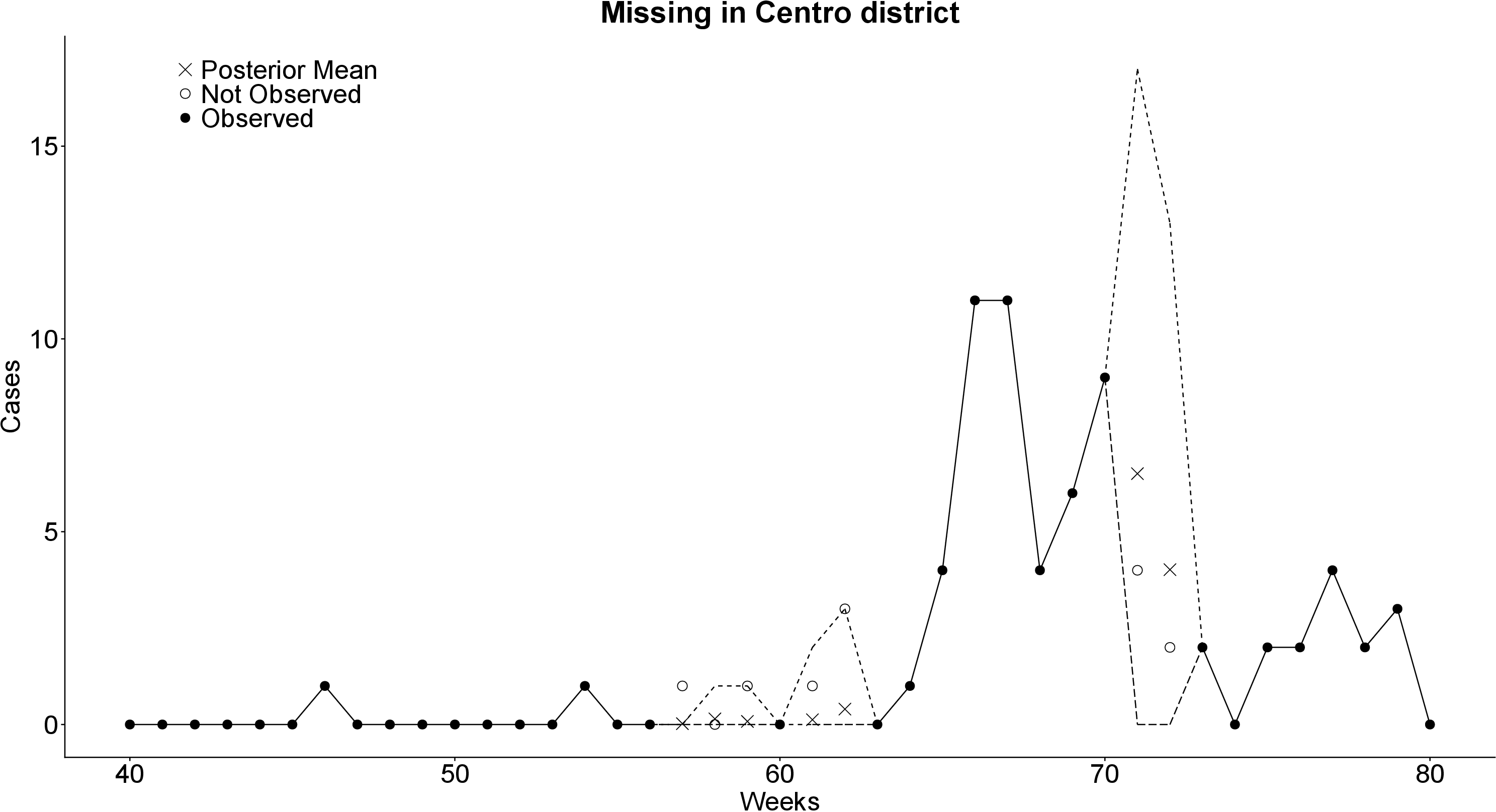}
    \caption{{First 40-80 weeks of observed reported cases in Centro district (denoted as black solid circles), with the reported case at the 57th-59th, 61st-62nd and 71st-72nd weeks removed artificially (denoted as open circles). Crosses represent the posterior mean from the proposed ZS-MSNBH model and dashed lines represent the  95\% predictive credible intervals.} }
    \label{fig:missing-in-mid}
\end{figure}}

\newpage
\bibliography{supplement}
